\documentclass{aa}

\usepackage{graphicx}
\usepackage{txfonts}
\usepackage{lscape}
\newcommand{\kms}{$\mathrm{\,km\,s}^{-1}$ }

\begin{document}

\title{Vertical distribution of Galactic disk stars
\thanks{Based on observations  made at the Observatoire
        de  Haute  Provence (OHP, France). Data only available in electronic 
	form at the CDS (Strasbourg, France)}
\thanks{Full Tables \ref{t:TGMET_lib} and \ref{t:full} are only available electronically at the CDS}} 
\subtitle{IV - AMR and AVR from clump giants } 

\titlerunning{AMR and AVR of the thin disk}
\author{C. Soubiran\inst{1}, O. Bienaym\'e\inst{2}, T.V. Mishenina\inst{3}, 
V.V. Kovtyukh\inst{3}}
\authorrunning{C. Soubiran et al.}
\offprints{soubiran@obs.u-bordeaux1.fr}
%
\institute{Universit\'e Bordeaux 1 - CNRS - Laboratoire d'Astrophysique de Bordeaux, BP 89,  33270
              Floirac, France  \and Universit\'e de Strasbourg, CNRS Observatoire Astronomique, 11 rue de l'Universit\'e, 67000 Strasbourg, France
	      \and Astronomical Observatory of Odessa
National University, Shevchenko Park, 65014, Odessa, Ukraine}
\date{Received : October 4, 2007 / Accepted : November 30, 2007 }
\abstract{We present the parameters of 891 stars, mostly clump giants, including atmospheric parameters, 
distances, absolute magnitudes, spatial velocities, galactic orbits and ages. One part of this sample consists of 
local giants, within 100 pc, with atmospheric parameters either estimated from our spectroscopic 
observations at high resolution and high signal-to-noise ratio, or retrieved from the literature.  
The other part of the sample includes 523 distant stars,
spanning distances up to 1 kpc in the direction of the North Galactic Pole, for which we have estimated 
atmospheric parameters 
from high resolution but low signal-to-noise Echelle spectra. This new sample is  
kinematically unbiased, with well-defined boundaries in magnitude and colours. We revisit the 
basic properties of the 
Galactic thin disk as traced by
clump giants. We find the metallicity distribution to be different from that of dwarfs, with less metal-rich stars. 
We find evidence for a vertical metallicity gradient of -0.31 dex kpc$^{-1}$ and for a transition  at 
$\sim$ 4-5 Gyr
in both the metallicity and velocities. The age - metallicity relation (AMR), which exhibits a very
low dispersion, increases smoothly from 10 to 4 Gyr, with a steeper increase for younger stars. The 
age-velocity relation (AVR) is characterized by the saturation of the $V$ and $W$ dispersions at 5 Gyr, and
continuous heating in $U$.

\keywords{Stars: kinematics -- Stars: fundamental parameters -- Galaxy: disk --
Galaxy: structure -- Solar neighbourhood}
}
\maketitle
%
%
\section{Introduction}
This paper is the continuation of previous papers (Soubiran et al. \cite{sou03}, hereafter
Paper I and  
Siebert et al. \cite{sie03}, hereafter Paper II)
where we investigated the vertical distribution of disk stars with local and distant samples 
of clump giants. Our main result in Paper I was a new characterization of the thick disk, showing 
a rotational lag  of $-51  \pm  5\,\mathrm{km\,s}^{-1}$ with
respect  to  the Sun,  a  velocity  ellipsoid of $(\sigma_U,  \sigma_V,
\sigma_W)=(63\pm  6, 39\pm  4, 39\pm  4)  \,\mathrm{km\,s}^{-1}$, a mean
metallicity   of   [Fe/H]  =$-0.48\,\pm$\,0.05   and   a  high   local
normalization  of 15\,$\pm$\,7\%. We have also determined in Paper II the gravitational force 
perpendicular to the galactic plane  and  the mass density in the galactic plane 
($\Sigma = 67 M_\odot{\rm pc}^{-2}$) and thickness of the disk ($390^{+330}_{-120}$ pc). 
We found no vertex deviation for old stars, consistent with an axisymetric Galaxy. 
After these two papers, we have enlarged and improved our samples in order to go further into 
the study of the 
local thin disk. We have observed a large sample of  
local Hipparcos clump giants at high spectral resolution and high signal-to-noise ratio, and measured 
their metallicity and elemental abundances (Mishenina et al.
\cite{mish06}). Combined 
with a compilation of other studies providing metallicities of nearby clump giants, we have 
built a large unbiased sample of local giants to investigate the kinematical and chemical 
distributions 
of these stars. Our previous sample of distant giants was based on high resolution, low 
signal-to-noise spectra for 387 stars, spanning distances up to z=800 pc above the 
galactic plane, in the direction of the North Galactic Pole (NGP). The new distant sample now includes 
523 stars up to z=1 kpc, with improved distance and metallicity determinations.

These two improved samples, local and distant, have also been used for other purposes, presented in separate papers. Kovtyukh
et al. (\cite{kov06}) use the local sample to establish an accurate temperature scale for 
giants using line-depth ratios. Mishenina
et al. (\cite{mish06}) investigate mixing processes in the atmosphere of clump giants. 
Finally Bienaym\'e et al. (\cite{bie05}), hereafter Paper III, apply two-parameter models 
on the combination
of the local and distant samples to derive a realistic
estimate of the total surface mass density within 0.8 kpc and 1.1 kpc from the Galactic plane, 
respectively  $\Sigma_{\rm0.8\,kpc}$\,=\,59-67 $\,\mathrm{M}_{\sun}\mathrm{pc}^{-2}$ and 
$\Sigma_{\rm1.1\,kpc}$\,=\,59-77 $\,\mathrm{M}_{\sun}\mathrm{pc}^{-2}$. 

Here we use these new data to focus on local properties of the thin disk which are so important to
constrain its chemical and dynamical evolution : metallicity distribution,
vertical metallicity gradient, age - metallicity relation (AMR) and age - velocity relation (AVR). Numerous studies of these
properties have been published, with however considerable disagreements reflecting the variety of tracers (open clusters,
planetary nebulae, field dwarfs), discrepant metallicity scales, different age determinations, or selection biases.
 A major contribution
on the subject comes from the Geneva-Copenhagen survey of 
the Solar neighbourhood by Nordstr\"om et al. (\cite{nor04}), which includes stellar parameters similar 
to ours, but for a much larger sample of dwarfs, and with photometric, less reliable, metallicities. In the present 
work, the use of
distant giants allows us to probe larger distances above the galactic plane where kinematical 
distributions are no longer affected by local streams and moving groups, as studied by Famaey et al. 
(\cite{fam05}). Moreover, giants are well suited for age determinations, as shown in da Silva et al.
(\cite{dasil06}). We use their Bayesian method with isochrone fitting to compute ages and, similarly to them,
we use the complete resulting probability distribution function of each star to bin the age axis. The
combination of this pertinent method with the fact that we use spectrocopic metallicities for a large,
homogeneous and complete sample, with well defined boundaries in magnitude and colour, should ensure that 
the new relations that we obtain are quite reliable. We have also
computed for each star its probability of belonging, on kinematical criteria,  to the thin disk, the thick disk, 
the Hercules stream and the halo, in order to reject the most probable non thin disk stars.

Sections \ref{s:hip_sample} and \ref{s:pgn_sample} describe the local 
and distant samples. We give details on the TGMET method and the new reference library which have
been used to improve the determination of $T_{\rm eff}$, $\log\,g$, $\mathrm{[Fe/H]}$,
and $M_{\rm v}$ for the distant giants observed at high spectral resolution, but low signal-to-noise 
(Section \ref{s:TGMET}). Sections \ref{s:ages} and \ref{s:pop} describe the determination of ages, 
Galactic orbits and population membership. Then we select the most probable thin disk clump giants and 
demonstrate
the existence of a vertical metallicity gradient (Section \ref{s:grad}). We present the AMR derived from the
same stars in Section \ref{s:AMR}, while in Section \ref{s:AVR} we discuss the AVR in U, $V$ and $W$ derived
from a larger sample of clump giants where the most probable thick disk, Hercules stream and halo members
have been rejected.

%
\section{The local sample of Hipparcos giants}
\label{s:hip_sample}

The sample of local giants, dominated by clump giants, consists of the 381 single Hipparcos field stars 
which follow the criteria :
$$ \pi \ge 10\, \rm{mas}$$
$$ \delta_{ICRS} \ge -20^\circ$$
$$0.7 \le B-V \le 1.2$$  
$$ M_{\rm V} \le 1.6 $$

where $\pi$ is the Hipparcos parallax and $ \delta_{ICRS}$ the declination. It is thus
a complete sample.
The Johnson B-V colour has been obtained from the Tycho2 $B_{\rm  T}-V_{\rm
T}$ colour  applying Eq. 1.3.20 from ESA (\cite{esa97}) :
$$B-V = 0.850 \,(B{\rm _T}-V{\rm _T})$$

Absolute  magnitudes  $M_{\rm{v}}$  were computed with 
V apparent magnitudes resulting from the transformation of  Hipparcos magnitudes $H_{\rm p}$
to the Johnson system, calibrated by Harmanec (\cite{har98}). 

Radial velocities have been mainly compiled from observations on the ELODIE spectrograph at Observatoire de Haute-Provance (OHP). Some 
177 local giants have been observed for this project (Mishenina et al. \cite{mish06}), while radial 
velocities of other stars were retrieved from  
the ELODIE library
(Prugniel \& Soubiran \cite{pru01}, \cite{pru04}) and the ELODIE archive (Moultaka et al. \cite{mou04}).
For the remaining stars, we found radial velocities in Famaey et
al. (\cite{fam05}) and Barbier-Bossat et al. (\cite{bar00}). In summary, we have retrieved radial 
velocities for 220 stars in the various
ELODIE datasets, for 54 stars in Famaey et al's catalogue, for 107 stars in Barbier-Bossat 
et al's catalogue. We have also retrieved from these different sources information about the binarity
of the stars. We have flagged 30 suspected spectroscopic binaries presenting an enlarged or double peak
of their cross-correlation function.

Atmospheric parameters ($T_{\rm eff}$, $\log\,g$, [Fe/H]) have been compiled from the [Fe/H] 
catalogue (Cayrel de Strobel et
al. \cite{cay01}) updated with a number of recent references. The [Fe/H] 
catalogue is a bibliographical compilation which lists determinations of atmospheric parameters relying on
high resolution, high signal-to-noise spectroscopic observations and published in the main astronomical
journals. We have added to the compilation effective temperatures determined  
by Alonso et al. (\cite{alo01}), di Benedetto (\cite{diben98}), 
Blackwell \& Lynas-Gray
 (\cite{blac98}) and Ram\'\i rez \& Mel\'endez (\cite{ram05}).  A number of other recent references providing
spectroscopic ($T_{\rm eff}$, $\log\,g$, [Fe/H]) have been added to the [Fe/H] catalogue in an effort to keep it
up to date. For the present work, the largest contributions come from 
Mishenina et al. (\cite{mish06}) for 177 stars and
da Silva et al. (\cite{dasil06}) for 14 stars. For the older references, which were already in Cayrel de Strobel et
al. (\cite{cay01}), the largest contribution comes from McWilliam (\cite{mcw90}) for 233 stars. 
This compilation provided [Fe/H] for 363 stars, adopting a weighted average when several values
where available for a given star (a higher weight was given to the most recent references). For
5 remaining stars, an ELODIE spectrum was available, enabling the determination of atmospheric
parameters with the TGMET method (see next section). We thus have just 13 stars which lack atmospheric
parameters, representing 3\% of the whole local sample.

Combining atmospheric parameters from different sources can be a source of errors if some verifications 
are not made. Not all authors of spectroscopic analyses use the same temperature scales, Fe lines, 
and atomic data so that
systematic differences may occur in the resulting metallicities. In the present work,  our narrow ranges in colour and luminosity
suggest we deal with a very 
limited 
range of atmospheric parameters where temperature determinations from different 
methods usually agree well. This is confirmed in our sample where 99 stars have at least two 
different determinations of $T_{\rm eff}$. Computing the mean $T_{\rm eff}$  for each of these 99 stars, we find standard deviations
ranging from 0 to 140 K, with a median value of 40 K, which is below the commonly admitted external error
on effective temperatures ($\sim$ 50-80 K). Only 6 stars show $T_{\rm eff}$  determinations deviating by more 
than 100 K. Similar verifications were made on [Fe/H] :  
the median value of standard deviations around the mean for stars having at least two 
determinations is 0.09 dex. 

Hipparcos proper motions and parallaxes have  been combined with radial velocities
through the  equations of Johnson \& Soderblom (\cite{joh87}) to compute  the 3  velocity components
$(U,V,W)$ with  respect to the Sun (the $U$ axis points towards the Galactic Center).  

\begin{figure}[htbp]
\center
\includegraphics[width=8cm]{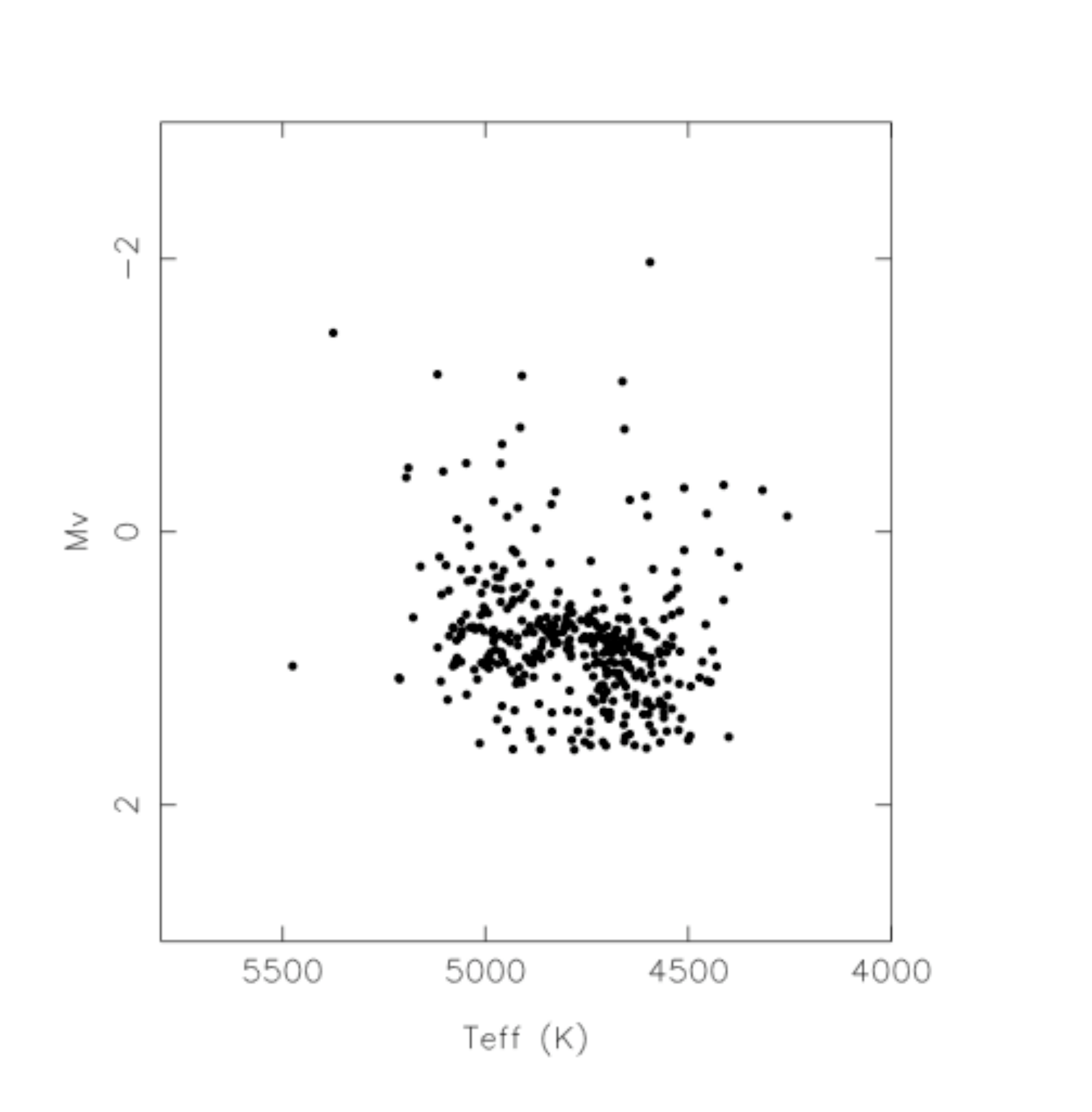}
\includegraphics[width=8cm]{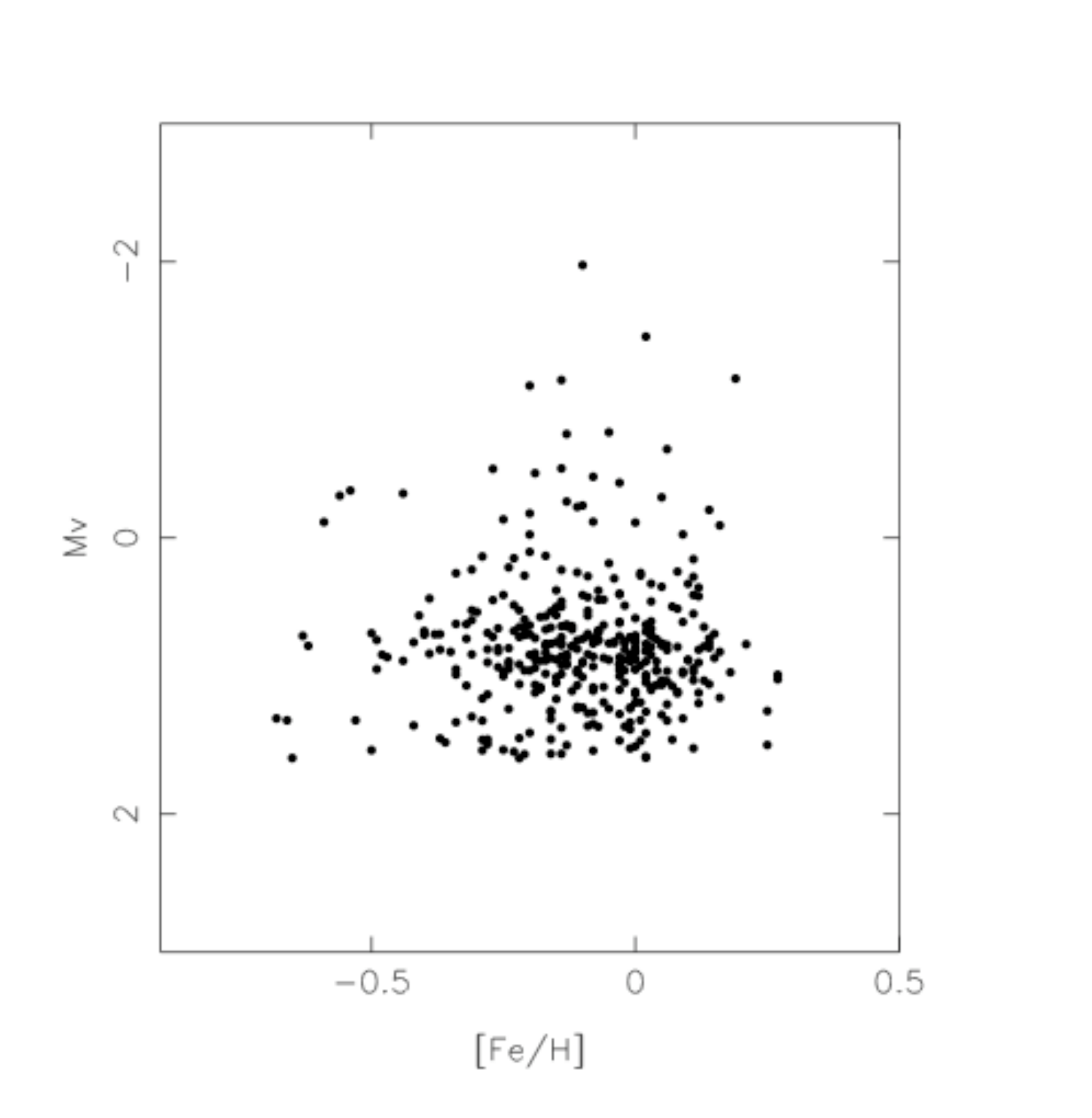}
\includegraphics[width=8cm]{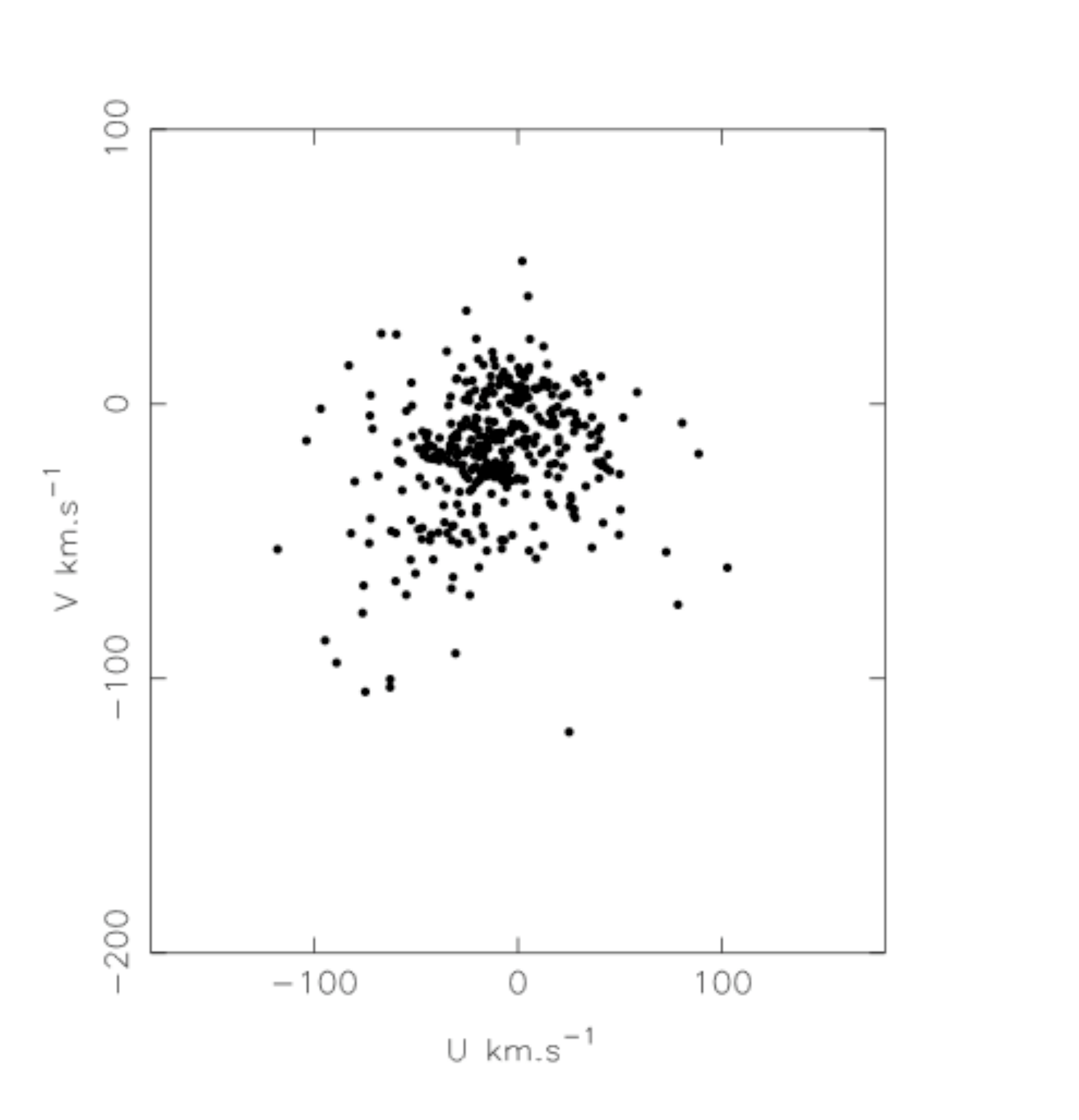}
\caption{Our local sample in the  $M_{\rm v}$ vs $T_{\rm eff}$,  $M_{\rm v}$ vs [Fe/H] and $V$ vs $U$ diagrams}
\label{f:loc}
\end{figure}

Figure \ref{f:loc} shows the distribution of this sample in the planes  $M_{\rm v}$  vs $T_{\rm eff}$,  $M_{\rm v}$  vs 
[Fe/H] and $V$ vs $U$.

%
\section{The distant NGP sample}						     
\label{s:pgn_sample}

The distant sample has been drawn  from the Tycho2 catalogue
(H{\o}g et al. \cite{hog00}).  We have applied similar criteria as in
Soubiran  et  al.  (\cite{sou03})  to  build  the  list of  red  clump
candidates, just extending the limiting apparent magnitudes to fainter
stars. A detailed description of the sample can be found in Paper III. The resulting
sample consists of 523 different stars on
a 720 square degree field close to the NGP.
The Tycho2 catalogue provides accurate proper motions and $V$ magnitudes. 
High resolution spectroscopic observations on ELODIE allowed us to measure radial velocities,
spectroscopic distances and metallicities. 

\subsection{Spectroscopic observations, radial velocities}
The observations were carried out with the echelle spectrograph ELODIE
on the  1.93\,m-telescope at the  Observatoire de Haute  Provence. The
performances  of this  instrument  are described  in Baranne et al. (\cite{bar96}). Compared
to our previous study in Paper I, 141 additional spectra have been obtained in Febuary
and March 2003. The 
resulting 540 spectra  cover the full range
390 -- 680 nm at a resolving power  of 42\,000. The reduction has been made at the 
telescope with the on-line  software which performs the spectrum extraction,
wavelength calibration and measurement of radial velocities by cross-correlation with
a numerical mask.
 The radial velocity accuracy is better than
\mbox{$1\,\mathrm{km\,s}^{-1}$}  for the  considered stars  (K stars).
Our    sample   spans    radial   velocities    from   --139    to   85
$\mathrm{km\,s}^{-1}$    with     a    mean    value     of    $-12.8$
$\mathrm{km\,s}^{-1}$. The mean S/N of the spectra at 550 nm is 22. Some 17 stars have 
been observed twice. For 13 stars, the 
correlation peak was enlarged or double, indicating the probable binarity of these stars
which were flagged. 

\subsection{Stellar parameters ($T_{\rm eff}$, $\log\,g$, $\mathrm{[Fe/H]}$, 
$M_{\rm v}$)}
\label{s:TGMET}

We have performed the determination of stellar parameters 
$T_{\rm eff}$, $\log\,g$, $\mathrm{[Fe/H]}$ and $M_{\rm v}$ from
ELODIE spectra using the code TGMET (Katz et al. \cite{kat98}), like in Paper I.
TGMET relies on the  comparison by minimum
distance of target spectra to a library of  stars with well known
parameters, also  observed with ELODIE (Soubiran  et al. \cite{sou98},
Prugniel \& Soubiran \cite{pru01}). As compared to Paper I, we have improved
the content of the TGMET library because we were aware that the quality of TGMET 
results are very dependent of the quality of the empirical library
which is used as reference. We present in this section the library that we built for the
present study dealing with clump giants. We also present
the tests which have been performed to assess the reliability of the TGMET parameters.

\begin{figure}[htbp]
\center
\includegraphics[width=8cm]{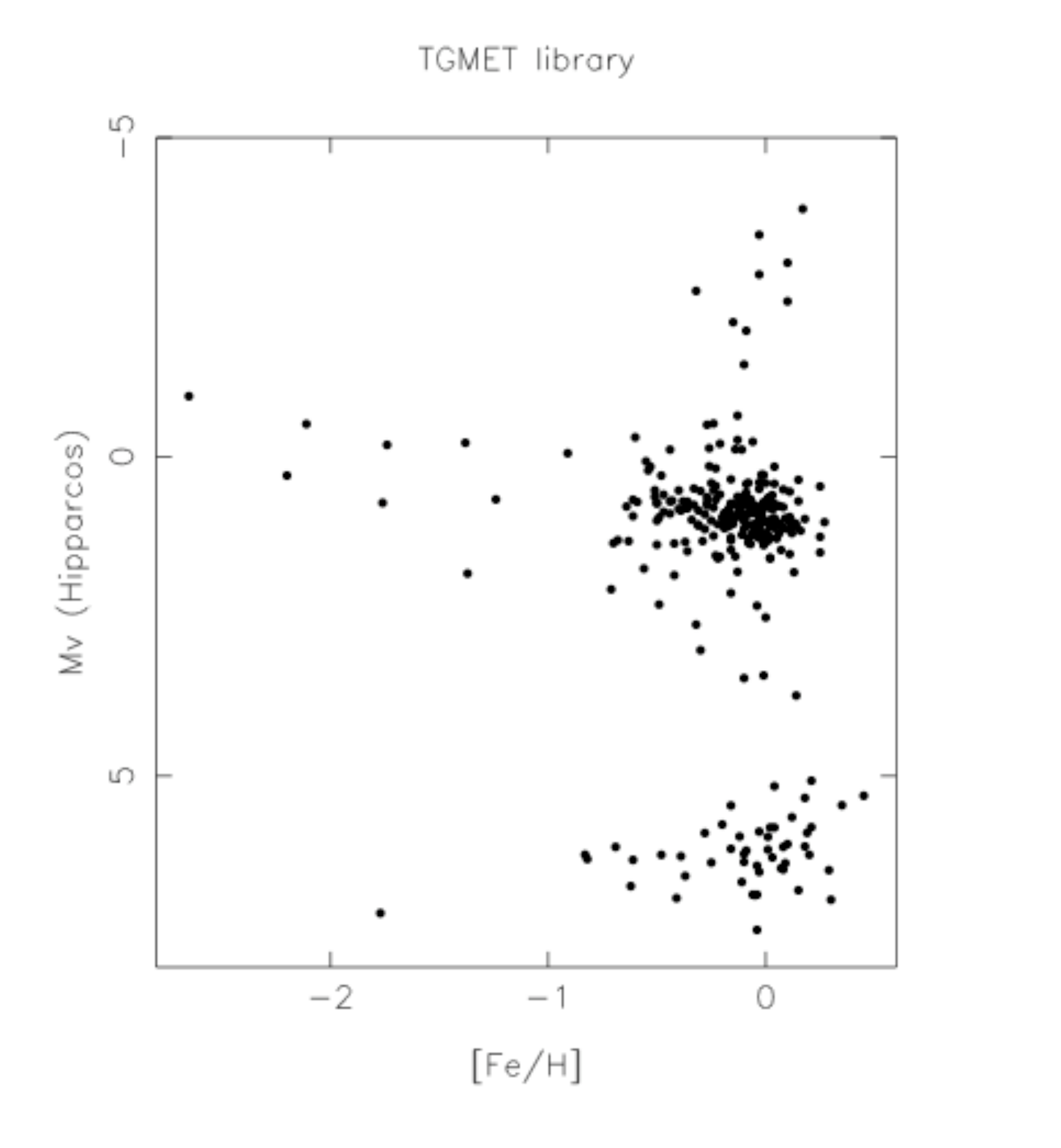}
\caption{Distribution of the TGMET library used in this study (724 reference 
stars observed with ELODIE) in the plane metallicity - absolute magnitude.}
\label{f:bib_feHMv}
\end{figure}

The TGMET library must be built with reference spectra representative of
the parameter space occupied by the target stars, with a coverage as dense
as possible. The parameters of the reference spectra must be known as accurately 
as possible. Since our previous study of clump
giants  at  the  NGP, in papers I and II,  the  TGMET  library  has  been  improved
considerably. Many  stars with well  determined atmospheric parameters, compiled from
the literature, 
and with accurate Hipparcos parallaxes,  have been added to the library
as   reference   stars  for   $T_{\rm   eff}$,   $\log\,g$,   [Fe/H] and
$M_{\rm{V}}$.  In particular the Hipparcos giants observed with ELODIE to build
the local sample and analysed by Mishenina et al. (\cite{mish06}) have been added to
the library.  Fig. \ref{f:bib_feHMv} shows
the distribution of the TGMET library used for this study in the
plane ([Fe/H], $M_{\rm{V}}$). The  clump area is densely covered down
to [Fe/H]  = $-0.80$. 

A small part of the TGMET library is presented in Table \ref{t:TGMET_lib}. The full Table is only
 available in electronic form, at the CDS. The calibrated Echelle spectra can be retreived from the ELODIE archive\footnote{http://atlas.obs-hp.fr/elodie/}. 

\begin{table*}
\caption{
Data and stellar parameters compiled for the TGMET library  : HD/BD number, date of observation,  $T_{\rm   eff}$,   $\log\,g$,   [Fe/H],
$M_{\rm{V}}$, quality flags qt, qf and qm for $T_{\rm   eff}$, [Fe/H] and $M_{\rm{V}}$ respectively (0 : data not available, 1 : poor, ... 4 : high), S/N of ELODIE spectrum at 550 nm, radial velocity, $B-V$, spectral type. The full table for all 724 stars is only available electronically at the CDS. The corresponding Echelle spectra can be retreived from the ELODIE archive.}
\label{t:TGMET_lib}
\centering
\begin{tabular}{l c c c c r c r r r l}
\hline
\hline
 HD/BD & date & $T_{\rm   eff}$ &   $\log\,g$ &   [Fe/H] & $M_{\rm{V}}$ & qt  qf qm & S/N & $RV$ & $B-V$ & ST \\
               &         &    K                     &                 &    dex    &                             &                &         & \kms &             & \\
\hline
 BD+430699   & 20040203  & 4760  & 4.68 & -0.41 &  6.916 & 424 & 130.6&     7.41&     0.972 &  K2   \\                            
 BD+522815   & 20040902  &           &       &  & 7.914 & 003  & 95.9&   -52.66&    1.164 & K5\\                               
 BD-004234   & 19970826  & 4574  & 4.32 & -0.84  & 6.237 & 233  & 79.2&  -127.58 &    0.968&  K3Ve+... \\                        
 HD001227    & 19970822  & 5037  & 2.65 & +0.25  & 0.465 & 422 & 101.1&    -0.04&    0.910 & G8II-III  \\                       
 HD002506    & 20001216  &            &     &   & 1.245 & 003  & 72.0&   -59.33&     0.933 & G4III  \\                          
 HD002910    & 20031101  & 4745  & 2.75 & +0.10  & 0.904 & 434 & 218.8&   -13.46&    1.074 & K0III    \\                        
 HD003546    & 19961003  & 4878  & 2.38 & -0.69  & 0.780 & 344 & 120.9&   -84.14&    0.843 & G5III... \\                        
 HD003651    & 20011125  & 5192  & 4.42 & +0.14  & 5.650 & 444 & 136.1&   -33.06&     0.849 & K0V   \\                           
 HD003712    & 19970822  & 4594  & 2.14 & -0.10 & -1.973 & 423 & 376.0 &   -4.49  &  1.182&  K0II-IIIvar  \\                    
 HD003765    & 19951030  & 5067  & 4.45 & +0.10  & 6.158 & 434 & 135.8 &  -63.32&    0.953 & K2V \\                             
 HD004188    & 20031101  & 4816  & 2.79 & +0.04  & 0.734 & 434 & 156.0&    -0.41 &   1.006 & K0IIIvar \\                        
 HD004256    & 20040903  & 4930  & 4.80 & +0.34  & 6.299 & 324 & 111.1 &    9.33 &   1.006 & K2V  \\                            
 HD004482    & 20021023  & 4917  & 2.65 & +0.02  & 0.991 & 424 & 151.7 &   -2.61 &    0.977 & G8II   \\                          
 HD004628    & 20010813  & 5040  & 4.64 & -0.25  & 6.360 & 444 & 147.1&   -10.35 &  0.876 & K2V \\                             
 HD004635    & 20031103  & 5129   &         &    & 6.072 & 303 & 155.1  & -31.75&    0.916 &  K0      \\     
 ... &    ... &... &... &... &... &... &... &... &... &...\\                
... &    ... &... &... &... &... &... &... &... &... &...\\  
\hline
\end{tabular}
\end{table*}

\begin{figure}[htbp]
\center
\includegraphics[width=8cm]{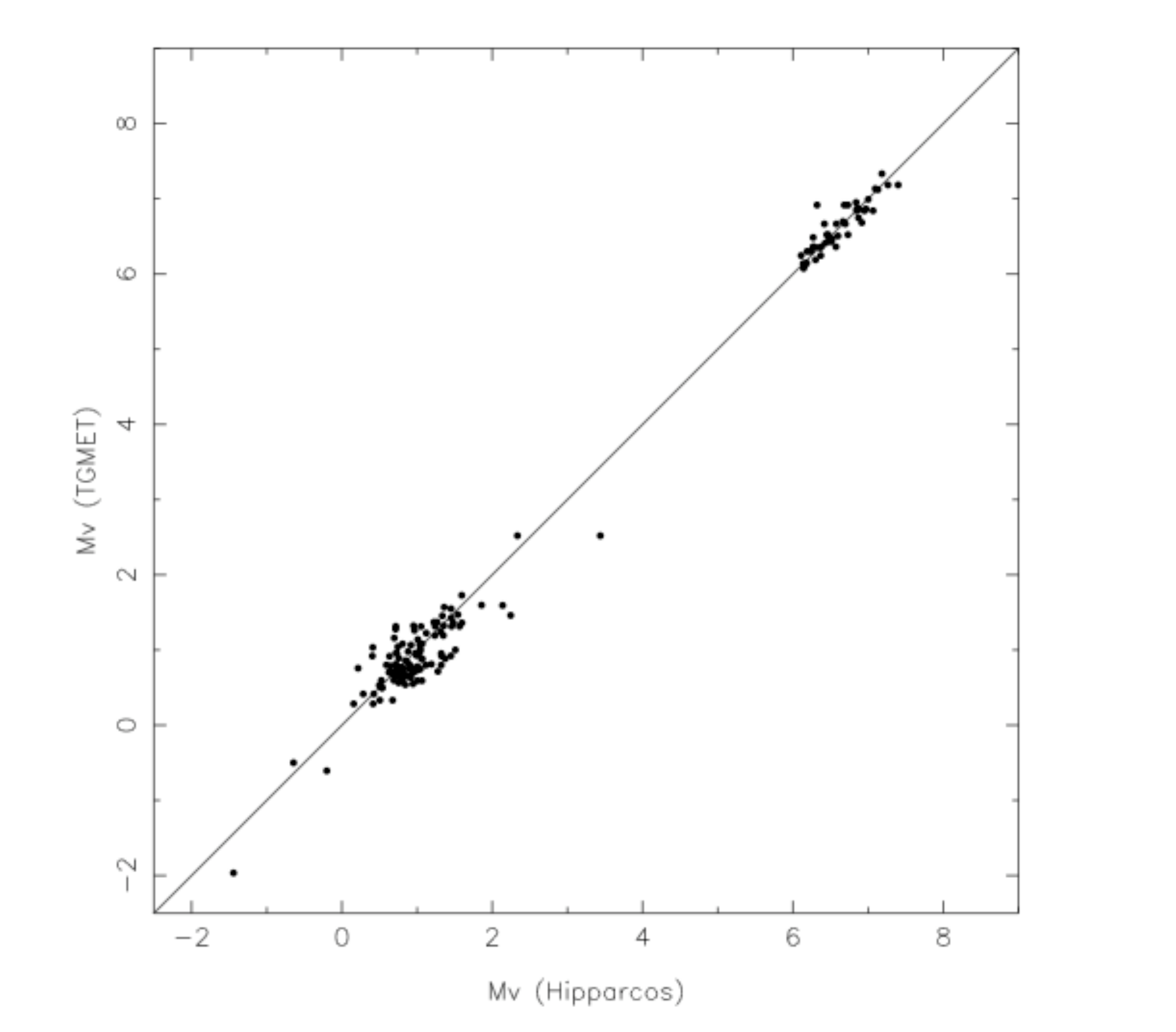}
\caption{Comparison of TGMET absolute magnitudes from degraded spectra 
to those deduced from
Hipparcos parallaxes for a subset of 158 reference stars.}
\label{f:compare_Mv}
\end{figure}

\begin{figure}[htbp]
\center
\includegraphics[width=8cm]{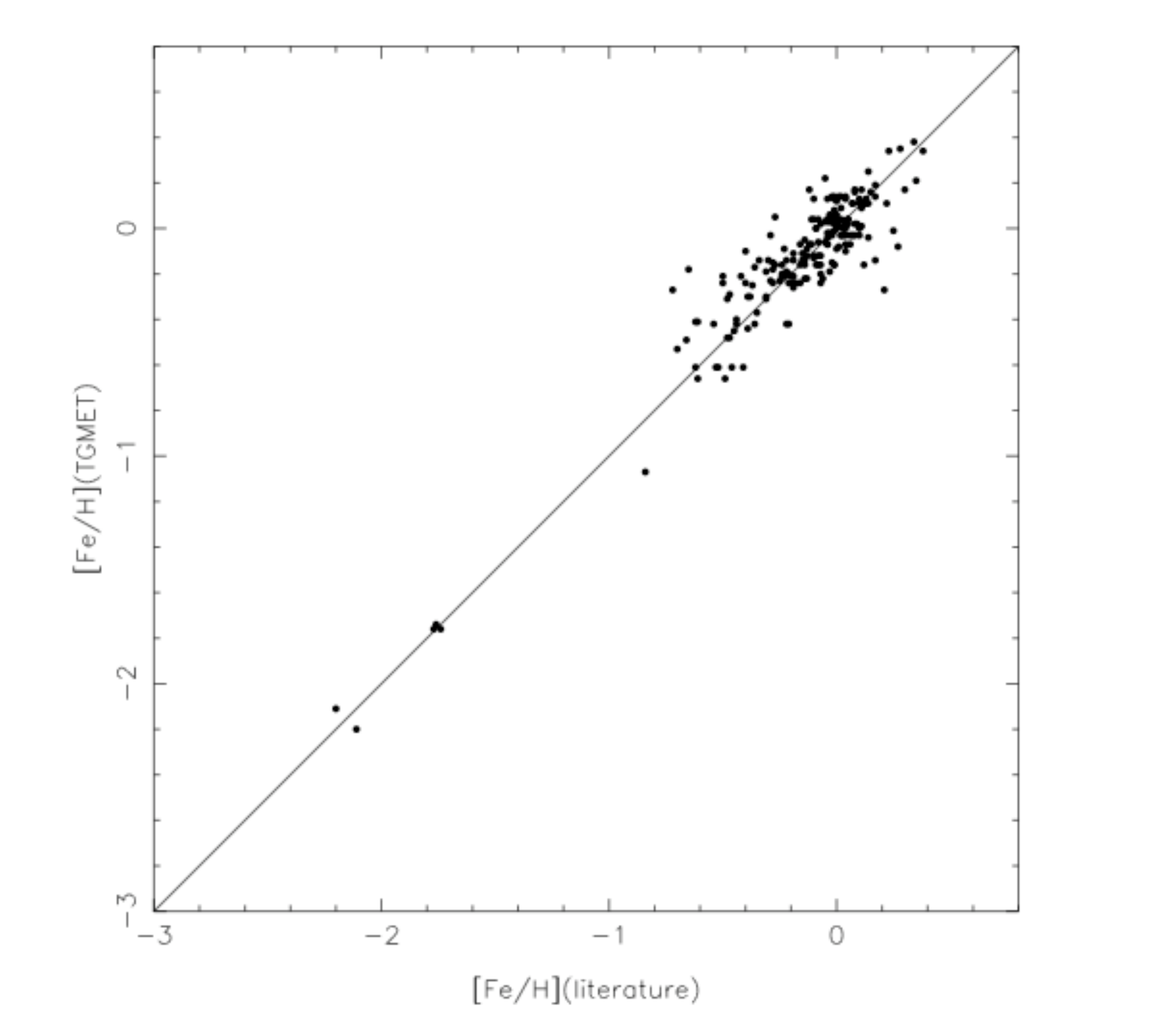}
\caption{Comparison of TGMET metallicities from degraded spectra to those from
the literature for a subset of 199 reference stars.}
\label{f:compare_FeH}
\end{figure}

In order to verify the TGMET results, we  have  run the code on ELODIE spectra
of stars chosen in the library, with the best known parameters, degraded to a S/N typical of our target
spectra (i.e. S/N=20). We have applied a bootstrap method : each test spectrum
was removed in turn from  the library, degraded to S/N=20,
 and its parameters determined
by comparison to the rest of the library.
To check results on
$M_{\rm{V}}$, we  have selected the 158  stars of the  library with a
relative error on their Hipparcos  parallax lower than 10\% and with 
$0.9 \le B - $V$ \le 1.1$. For  [Fe/H]  we have selected 199  stars  
with $0.9 \le B - $V$ \le 1.1$
having  the most reliable  spectroscopic
metallicity  determinations  found   in  the  literature. 
$M_{\rm{V}}$  and [Fe/H] determined from TGMET were  then compared  to their  Hipparcos and
literature  counterpart,  as  shown  in Figs.  \ref{f:compare_Mv}  and
\ref{f:compare_FeH}.    The  rms of the comparison,  respectively   0.25 mag  and   0.13 dex on
$M_{\rm{V}}$  and [Fe/H], measure  the accuracy  of the  TGMET
results at S/N=20. The rms on $M_{\rm{V}}$ corresponds to an error in distance
of 12\%.


\begin{figure}[htbp]
\center
\includegraphics[width=8cm]{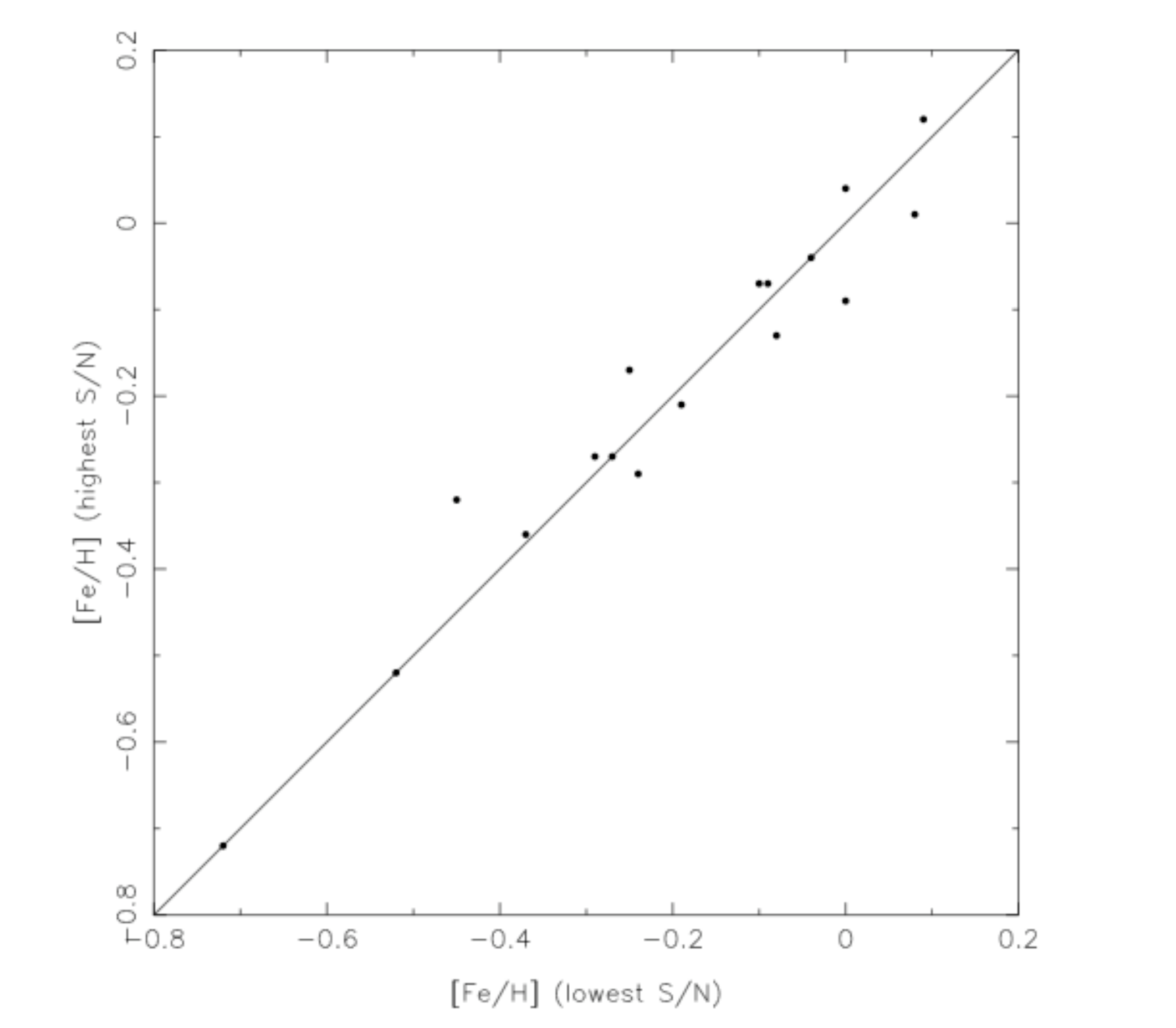}
\caption{Comparison of the TGMET metallicities obtained for the 17
target stars observed twice (rms=0.05 dex).}
\label{f:internal_FeH}
\end{figure}

In order to  test the internal precision of TGMET on [Fe/H], we
have compared  the results obtained  for the 17 stars observed
twice (Fig.~\ref{f:internal_FeH}).  As can  be seen, the  agreement is
very good (rms=0.05 dex).

\begin{figure}[htbp]
\center
\includegraphics[width=8cm]{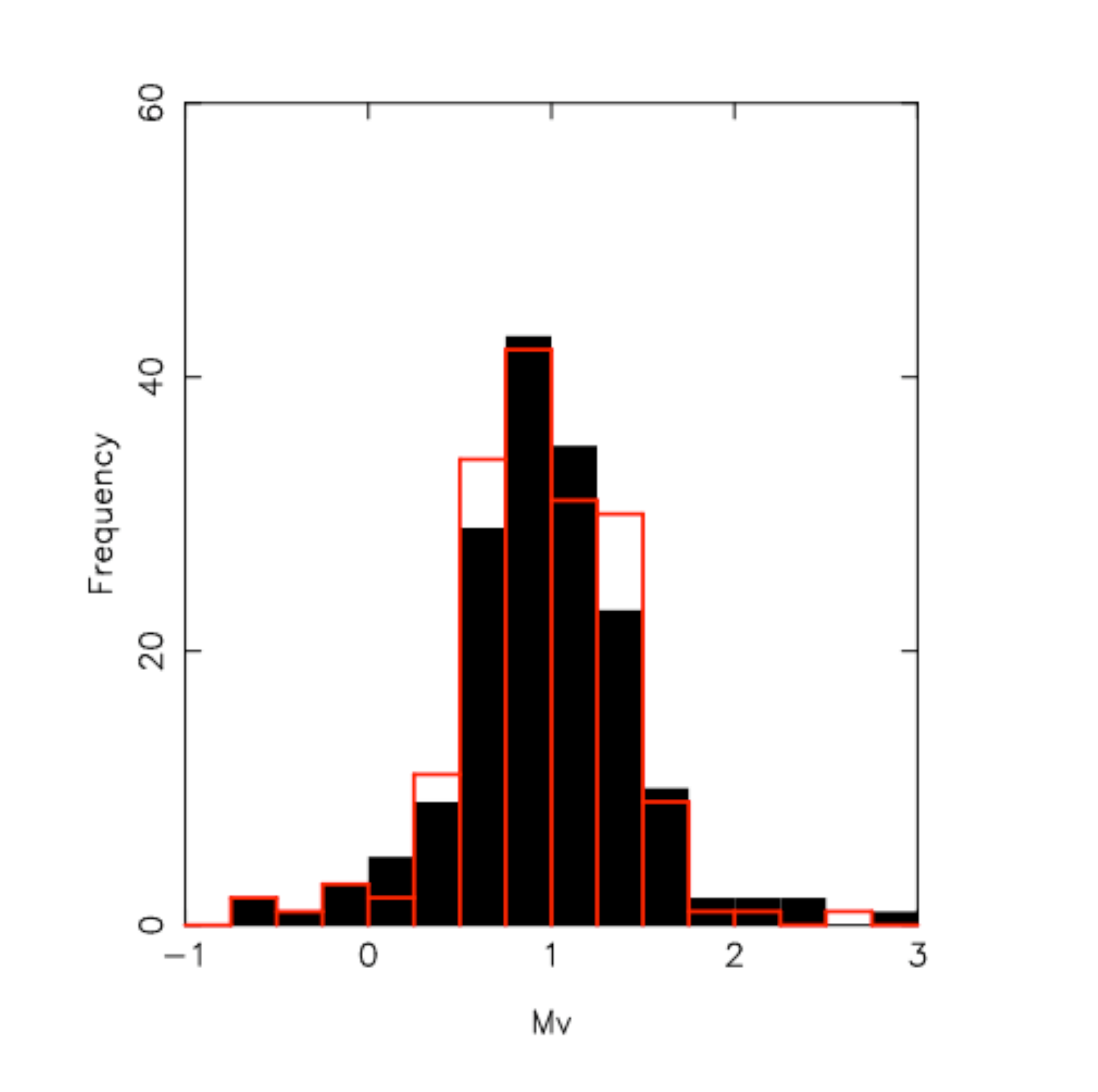}
\caption{Absolute magnitude histograms of the library's giants deduced  from
Hipparcos (filled) and deduced from the bootstrap test on degraded spectra (red line).}
\label{f:bib_Mv_histo}
\end{figure}

\begin{figure}[htbp]
\center
\includegraphics[width=8cm]{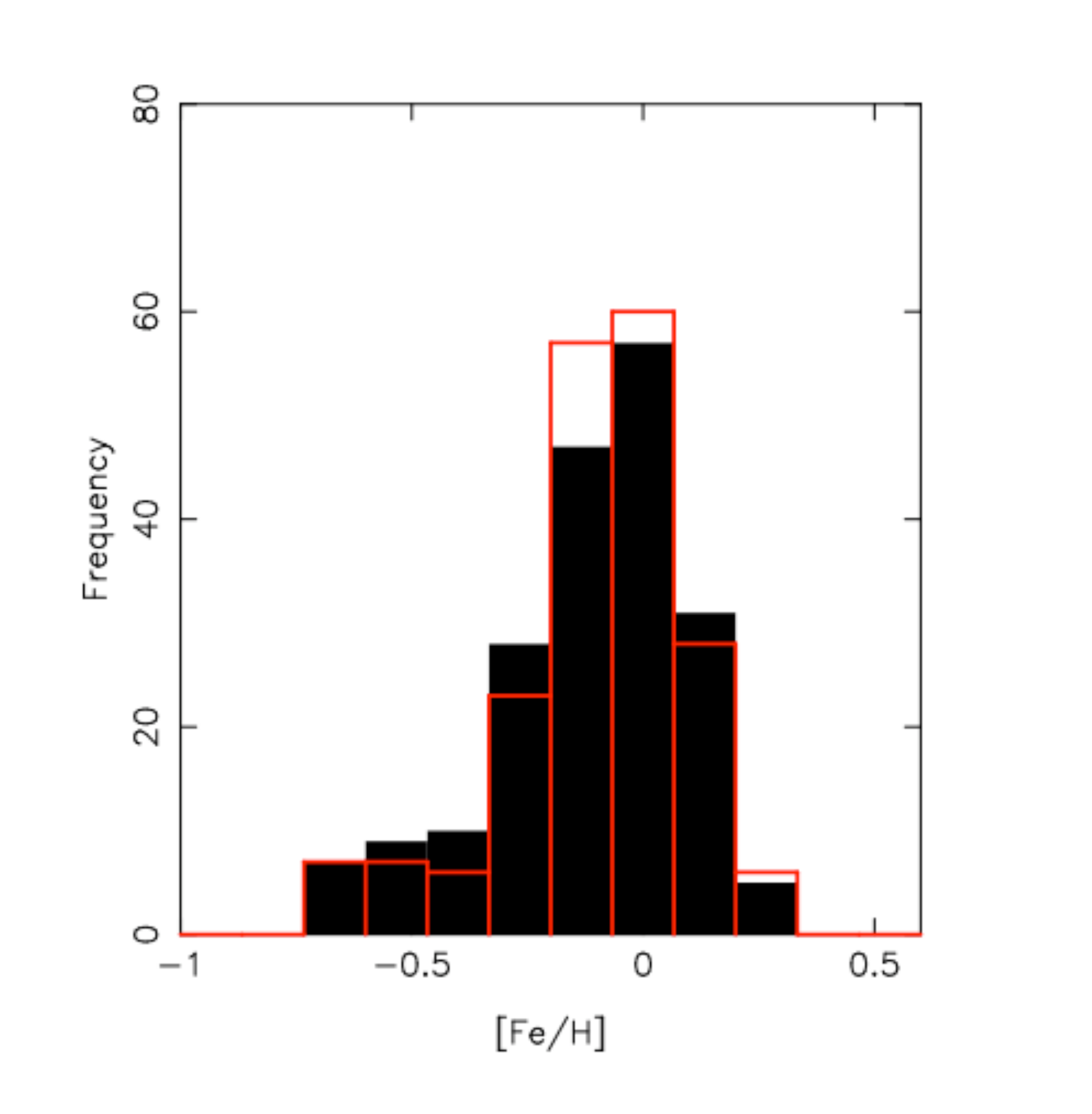}
\caption{Metallicity histograms of the library's clump giants deduced  from the
literature (filled) and deduced from the bootstrap test on degraded spectra (red line).}
\label{f:bib_feh_histo}
\end{figure}

An important verification has to be made to check that TGMET does not introduce a bias in the absolute
magnitude and metallicity
distributions of giants. In the following sections, parameters of distant giants, relying on TGMET, will be 
compared to parameters
of local giants, relying on literature and Hipparcos data. We thus have to ensure that these parameters are on 
the same scales.  Fig. \ref{f:bib_Mv_histo} shows the histograms 
of absolute magnitudes of the library's giants deduced from
Hipparcos and deduced from the bootstrap test on degraded spectra, in 0.25 mag bins. 
Similarly, Fig.~\ref{f:bib_feh_histo} shows the two
metallicity histograms, from  the
literature and from the bootstrap test. These histograms are
perfectly aligned and present similar dispersions which guarantees the lack of bias in the TGMET results.

\subsection{Distances, spatial velocities}
Distances  have been  computed  for  all  the target  stars  from  the  TGMET
$M_{\textrm  v}$ and  Tycho2  $V_T$  magnitude transformed  into  Johnson $V$.  No
correction of interstellar  absorption was applied since it  is supposed to be
very  low in  the NGP  direction.
Proper motions,  distances and  radial velocities  have  been combined
to compute  the 3  velocity components
$(U,V,W)$ with  respect to the Sun.  

\begin{figure}[htbp]
\center
\includegraphics[width=8cm]{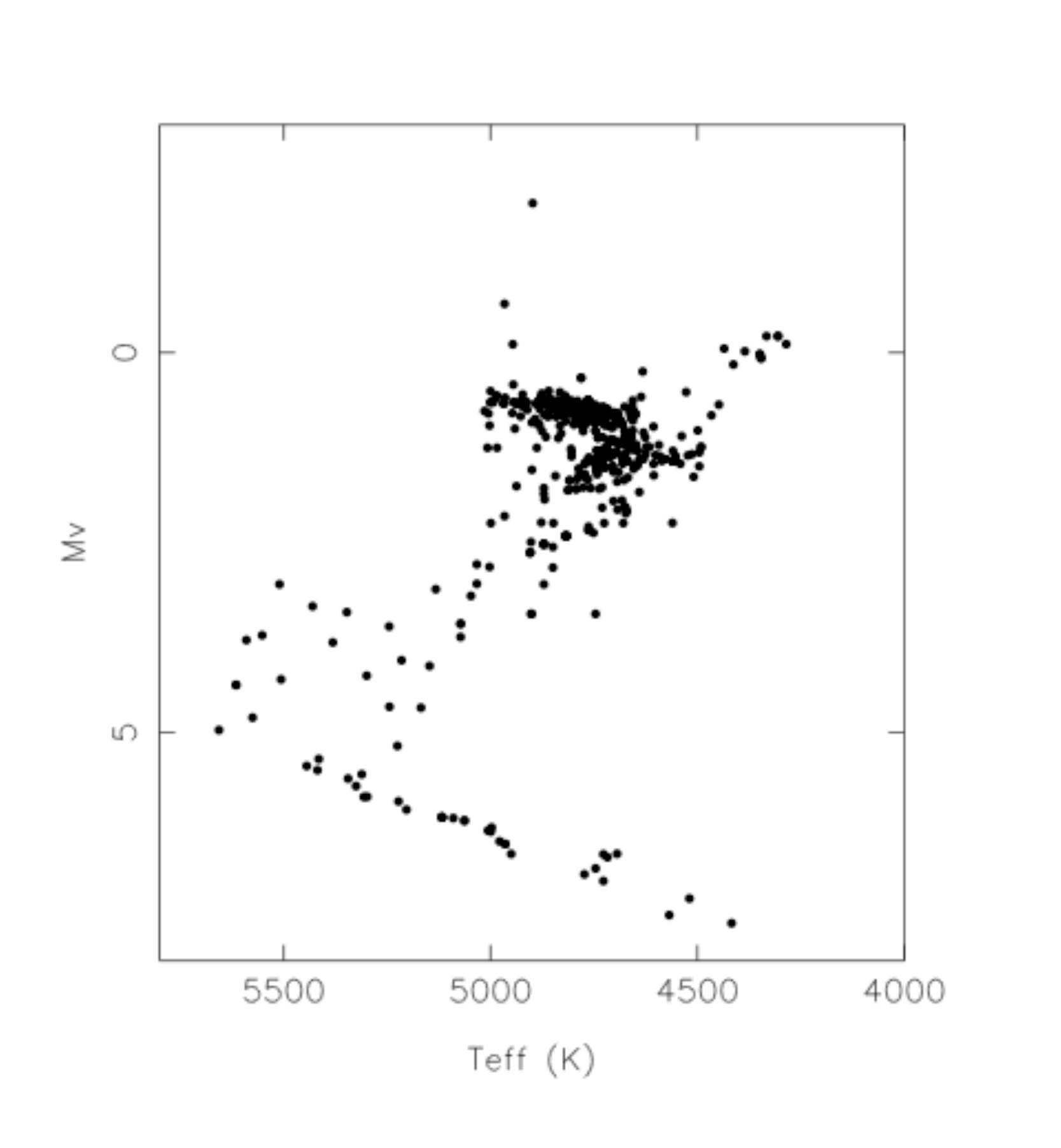}
\includegraphics[width=8cm]{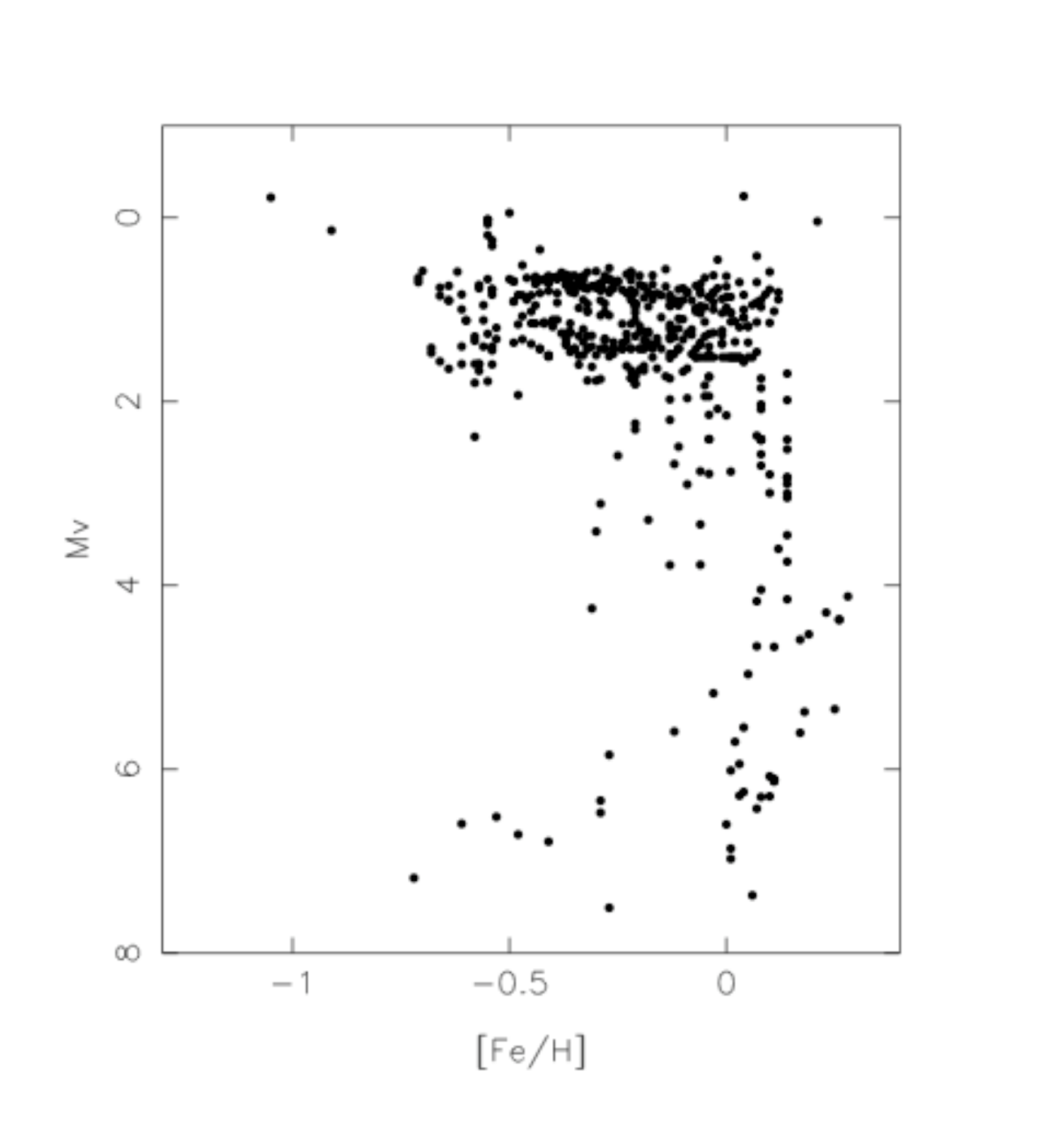}
\includegraphics[width=8cm]{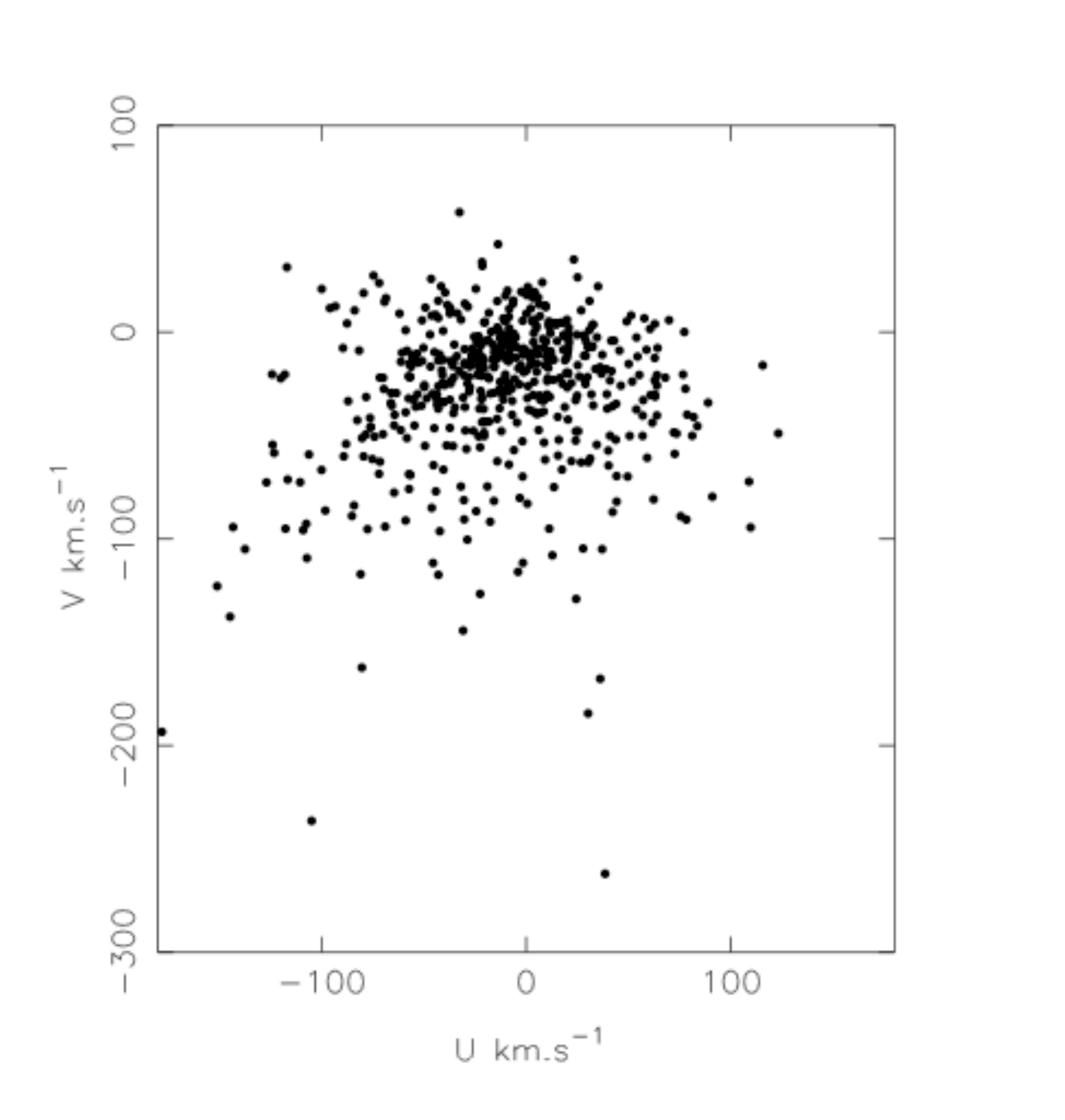}
\caption{The NGP sample in the  $M_{\rm v}$ vs $T_{\rm eff}$,  $M_{\rm v}$ vs [Fe/H] and $V$ vs $U$ diagrams.}
\label{f:pgn}
\end{figure}

Figure \ref{f:pgn} shows the distribution of  the 523 target stars in the planes  $M_{\rm v}$ vs $T_{\rm eff}$,  $M_{\rm v}$ vs 
[Fe/H] and $V$ vs $U$.

\section{Ages, Galactic orbits}
\label{s:ages}
\begin{figure}[htbp]
\center
\caption{Age - metallicity diagram for the 891 stars. Stars (83 local, 60 distant) with
relative age errors lower than 25 \% are highlighted as large filled circles.}
\includegraphics[width=8cm]{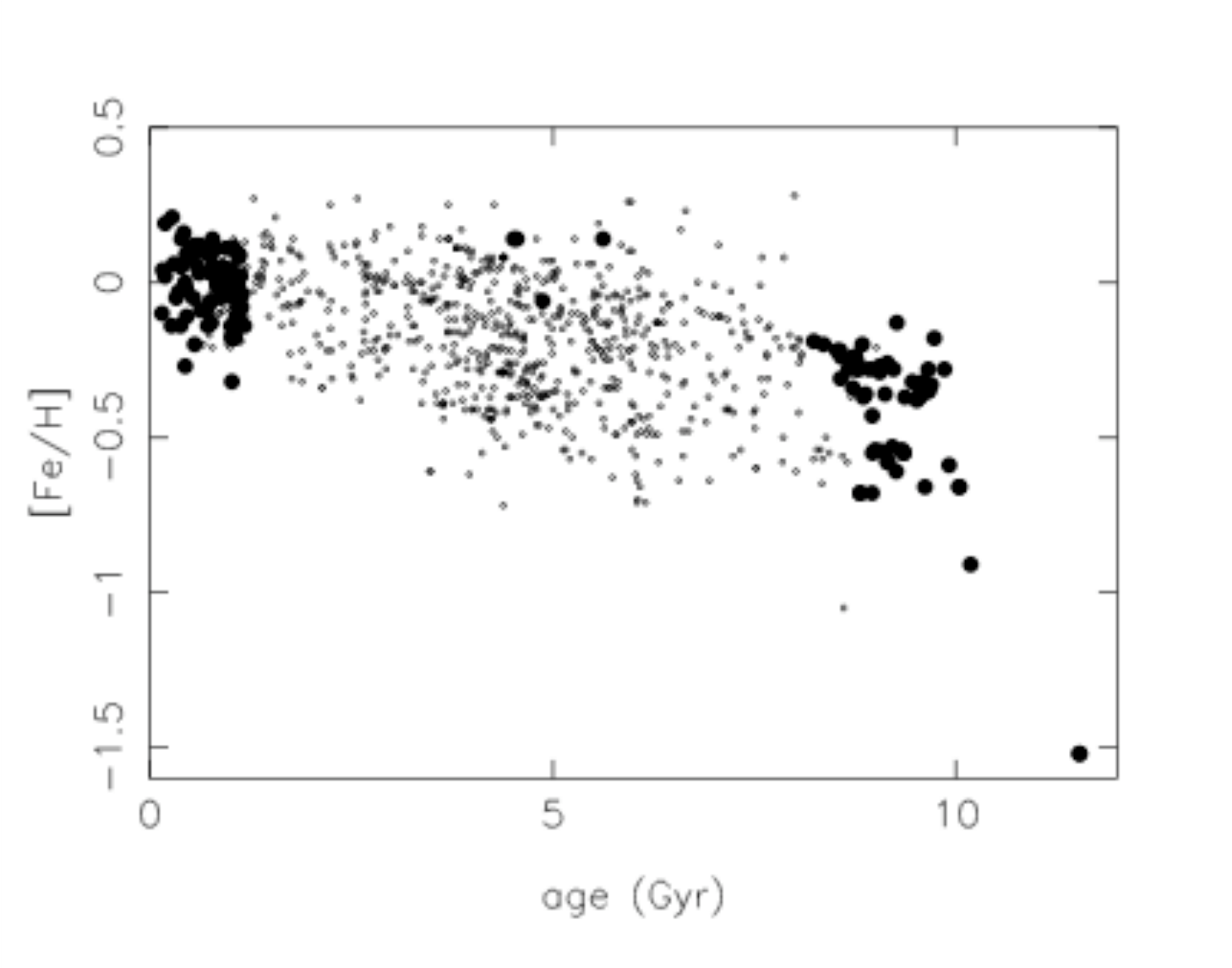}
\label{f:ld_af}
\end{figure}

Ages have been computed with the code PARAM developed by L. Girardi, available via an interactive 
web form\footnote{http://stev.oapd.inaf.it/~lgirardi/cgi-bin/param}. The method was initially developed by
J{\o}rgenson \& Lindegren (\cite{jor05}) and slightly modified as described in da 
Silva et al. (\cite{dasil06}). It is a Bayesian estimation method which uses theoretical isochrones computed by Girardi
et al. (\cite{gir00}) taking into account mass loss along the red giant branch. A convincing application of the
method to derive the fundamental parameters of evolved stars in an open cluster is presented in Biazzo et al. (\cite{bia07}).
Inputs to be given to the code are the observed effective temperatures, absolute 
magnitudes, metallicities and related errors. The output for each star is a probability
distribution function (PDF) of the age (and other parameters which are not used here). As shown in 
da Silva et al. (\cite{dasil06}), in their Fig. 5, the PDF of ages can be asymetric or even double 
peaked, especially in the case of red clump giants. As a consequence, ages are accurate for only
a tiny part of our sample. This should be kept in mind for the use of individual ages. Nevertheless the 
ages have significance when used statistically. As a proof, the age-metallicity plot for the 891 stars 
(Fig. \ref{f:ld_af}) shows a 
regular trend and a remarkably low dispersion as compared to other studies (e.g. 
Nordstr\"om et al. \cite{nor04}, da Silva et al. \cite{dasil06}). The 143 stars (83 local, 60 distant) with
relative age errors $<$ 25\% have been highlighted in Fig. \ref{f:ld_af}. Considering only 
these stars, we measure a mean metallicity of -0.06 with a dispersion of 0.10 dex for stars younger than 
2 Gyr, whereas the mean metallicity of older stars (age $>$ 8 Gyr) is -0.44 with a dispersion of
0.27 dex. There is no young star with a metallicity lower than -0.32, and no old star with
a metallicity higher than -0.13, contrary to common findings in samples of dwarfs, as for instance in Feltzing et al. (\cite{fel01}) and
 Nordstr\"{o}m et al. (\cite{nor04}). It is important to note this property of our sample, because the existence of old metal-rich stars is often mentioned to explain the large dispersion of the AMR (Haywood \cite{hay06}).  We come back to the AMR of the thin disk in Sect. \ref{s:clump}.

The orbital parameters have been computed by integrating the
equations of motion in the
galactic model of Allen \& Santillan (\cite{allen}), adopting a default
value of 4 Gyr as the integration time. The adopted velocity of the Sun with respect 
to the LSR is (9.7, 5.2, 6.7)
\kms (Bienaym\'e 1999), the solar galactocentric 
distance  ${\mathrm R}_{\odot}=8.5$ kpc
and circular velocity ${V_{\rm LSR}}=220$ \kms. 

\section{Population membership}
\label{s:pop}

The $U$ vs $V$ velocity distributions  of the local and distant samples can be compared from Figures~\ref{f:loc} and \ref{f:pgn}.
It is clear, from these plots, that the two samples contain different kinematical
populations. In the local sample, the velocities are clumpy and reflect 
moving groups and superclusters that dominate the kinematics in the solar neighbourhood.
Compared to Fig. 9 of Famaey et al. (\cite{fam05}), we can identify the Hercules stream at
$(U,V)\simeq (-40,-50)$ \kms, 
the Hyades-Pleiades supercluster at $(U,V)\simeq (-30,-20)$ \kms, and the Sirius 
moving group at $(U,V)\simeq (0,0)$\kms. There are very few high 
velocity stars that could correspond to the thick disk.
On the contrary, the velocities of the distant sample are better mixed with 
higher dispersions. This reflects the dynamical heating of the disk together with the growing number 
of thick stars with
increasing distance to the plane. In order to build a sample of pure thin disk
stars, we have performed the classification of all the stars into different 
kinematical populations. We have taken into account the Hercules stream because its 
velocity ellipsoid is just
intermediate between that of the thin disk and the thick disk, and is likely to
contaminate both populations. We did not attempt to distinguish the other
groups of the thin disk.\\

We assign to each star its probability to belong to the thin disk, the thick disk, the Hercules 
stream and the halo on the basis of its $(U,V,W)$ velocity and the velocity ellipsoids of 
these populations, in the same way as Soubiran \& Girard (2005) and with similar kinematical
parameters of the populations. In the distant sample we find
that 305 stars and 65 stars
have a probability higher than 80\% to belong to the thin disk and the thick disk
respectively. In the local sample, the numbers are 304 and 11.

\begin{figure}[htbp]
\center
\caption{Age - metallicity diagram for stars with
relative age errors lower than 25 \% and belonging to the thin disk (crosses) and
the thick disk (filled circles).}
\includegraphics[width=8cm]{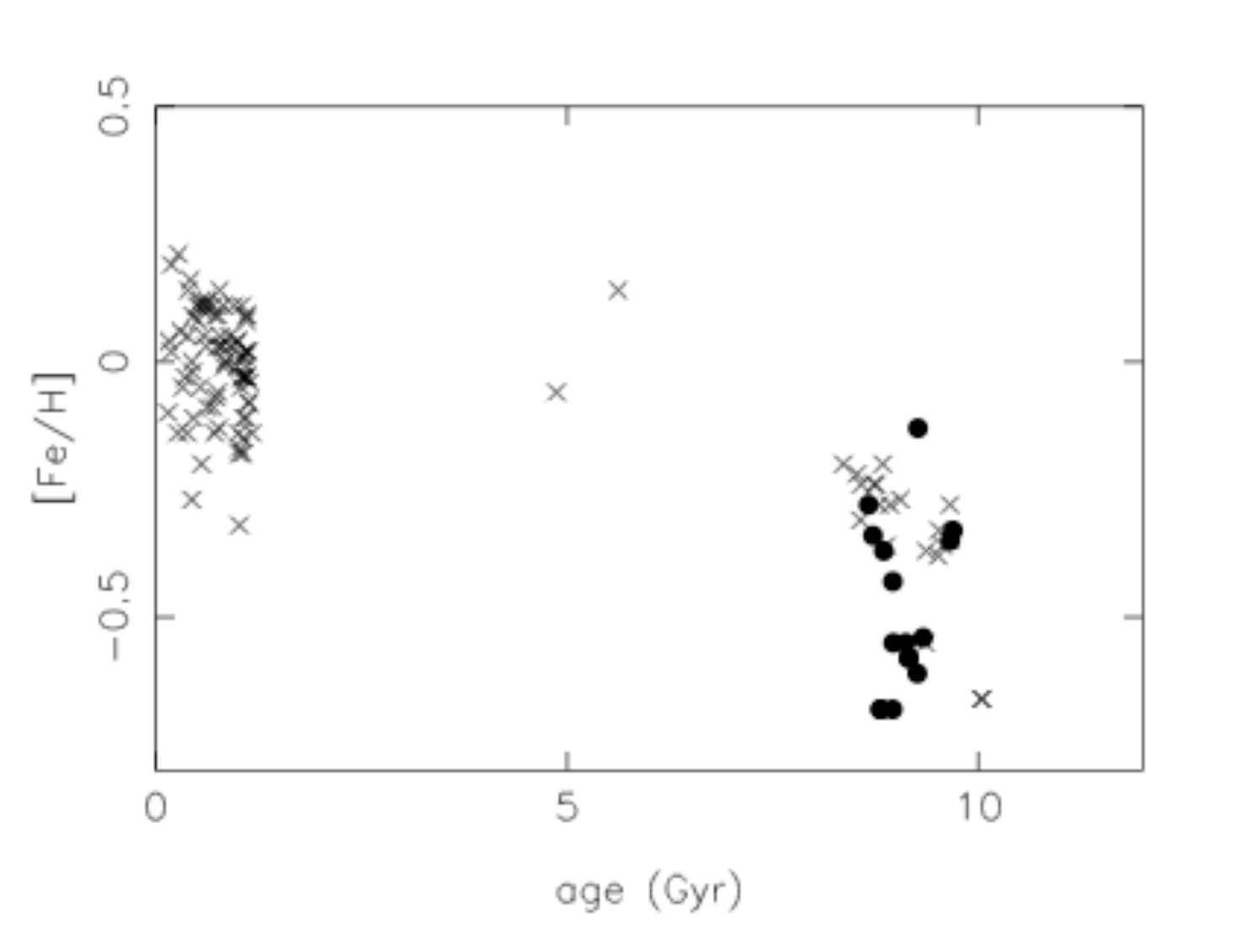}
\label{f:tt_af}
\end{figure}

One important question that we can immediately investigate thanks to this kinematical classification 
is whether the thin disk and the thick disk overlap in age and metallicity. 
Our data strongly suggest that this is the case.
Fig. \ref{f:tt_af} shows with different symbols the age-metallicity diagram for the most probable 
thin disk and the thick disk stars, restricted to relative age errors lower than 25 \% 
(suspected binaries rejected). It is clear 
that the oldest 
thin disk stars and thick disk stars overlap in the metallicity range -0.30 $\leq$ [Fe/H] $\leq$ -0.70, and age range 8-10 Gyr. It is also 
worth noticing that  there are no young thick disk stars.

All the parameters which have been determined as described in the previous sections are presented 
in Table \ref{t:full}. The full table with  all  891 stars is only available in electronic form at the CDS. The
file with the age PDFs is also available upon request. 

\begin{landscape}
\begin{table}
\caption{Stellar parameters of the programme stars derived in this work. The four columns p1, p2, p3 and p4 refer to the probability of belonging to  the thin disk, the thick disk, the Hercules stream and the halo respectively. SB=b indicate a suspected spectroscopic binary. }
\label{t:full}
\centering
\begin{tabular}{lcrrrrrrrrrrrrrrrrr} 
\hline\hline             
ID & $M_{\rm v}$ & $B-V$ & [Fe/H] & U & V & W & Rmin & Rmax & $|$Zmax$|$ &  ecc & d & age & $\sigma_{\rm age}$ & p1 & p2 & p3 & p4 & SB \\
    &              &                        & dex & \kms & \kms & \kms & kpc & kpc & kpc &  & pc & Gyr & Gyr &  & & &  & \\
\hline
 ..................... &  ......  & ......        &......  &......       &......      &......        &......  &......       &......    &......   &......   &......   &......      &......  &......  &......  &......  &  \\
 HD166229    &  1.471 &  1.165 &  0.01 &    50.3 &  -38.6 &    6.9 &   5.80 &   9.40 &   0.17 & 0.24 &   64 &  2.68 &  0.96 & 0.89 & 0.11 & 0.00 & 0.00 &    \\
HD169913    &  1.510 &  1.050 &  0.00 &   -25.4 &   -5.6 &    0.4 &   7.96 &   8.87 &   0.08 & 0.05 &  100 &  1.32 &  0.39 & 0.98 & 0.02 & 0.00 & 0.00 &    \\
 HD171994    &  1.549 &  0.882 & -0.23 &   -45.1 &  -15.9 &   -4.6 &   7.10 &   9.18 &   0.03 & 0.13 &   90 &  1.74 &  0.54 & 0.97 & 0.03 & 0.00 & 0.00 &    \\
 HD180610    &  1.525 &  1.160 & -0.01 &    14.7 &  -25.6 &  -10.4 &   6.90 &   8.74 &   0.05 & 0.12 &   50 &  7.01 &  2.02 & 0.96 & 0.04 & 0.00 & 0.00 &    \\
 HD192836    &  1.321 &  1.040 &  0.01 &     3.5 &   -9.2 &   -8.3 &   7.99 &   8.66 &   0.02 & 0.04 &   91 &  1.85 &  1.18 & 0.98 & 0.02 & 0.00 & 0.00 &    \\
 HD196134    &  1.564 &  0.984 & -0.14 &    25.1 &   -2.9 &  -17.9 &   7.77 &   9.58 &   0.14 & 0.10 &   97 &  4.46 &  1.94 & 0.98 & 0.02 & 0.00 & 0.00 &    \\
 HD198431    &  1.453 &  1.061 & -0.37 &   -52.3 &  -42.4 &  -20.3 &   5.76 &   8.94 &   0.17 & 0.22 &   77 &  9.55 &  1.69 & 0.35 & 0.08 & 0.57 & 0.00 &    \\
 HD211006    &  1.461 &  1.175 &  0.07 &   -22.1 &  -23.4 &   -8.7 &   7.11 &   8.59 &   0.04 & 0.09 &   77 &  4.88 &  1.79 & 0.96 & 0.03 & 0.01 & 0.00 &  \\
 HD212943    &  1.336 &  1.040 & -0.34 &    34.9 &  -16.1 &  -82.9 &   7.36 &   9.85 &   1.69 & 0.15 &   49 &  8.72 &  1.94 & 0.06 & 0.92 & 0.00 & 0.02 &    \\
 HD214995    &  1.363 &  1.101 & -0.09 &   -29.7 &  -36.6 &   -4.9 &   6.28 &   8.64 &   0.05 & 0.16 &   82 &  6.36 &  2.36 & 0.73 & 0.06 & 0.21 & 0.00 &    \\
 HD221833    &  1.585 &  1.155 &  0.02 &    24.0 &    3.5 &   -3.0 &   7.94 &   9.95 &   0.10 & 0.11 &   95 &  5.57 &  2.35 & 0.98 & 0.02 & 0.00 & 0.00 &   \\ 
 T0880-00075 &  0.693 &  0.918 & -0.49 &   -40.4 &  -33.9 &   81.8 &   7.05 &   9.10 &   2.21 & 0.13 &  423 &  4.62 &  2.72 & 0.02 & 0.97 & 0.00 & 0.01 &    \\
 T0880-00132 &  0.941 &  1.062 & -0.11 &   -30.6 &  -20.8 &  -22.9 &   7.27 &   8.72 &   0.35 & 0.09 &  266 &  3.93 &  1.67 & 0.92 & 0.07 & 0.01 & 0.00 &    \\
 T0880-00746 &  2.416 &  0.996 &  0.14 &     8.1 &  -33.2 &    3.8 &   6.51 &   8.62 &   0.24 & 0.14 &  197 &  3.71 &  1.14 & 0.90 & 0.07 & 0.03 & 0.00 &    \\
 T0881-00374 &  1.327 &  1.085 & -0.10 &     3.5 &  -38.4 &  -38.5 &   6.33 &   8.56 &   0.67 & 0.15 &  425 &  6.87 &  2.21 & 0.61 & 0.27 & 0.12 & 0.00 &   \\ 
 T0881-00435 &  0.816 &  1.000 & -0.27 &   -46.7 &  -16.7 &    6.3 &   7.18 &   9.20 &   0.37 & 0.12 &  316 &  4.69 &  2.70 & 0.95 & 0.05 & 0.00 & 0.00 &    \\
 T0881-00494 &  0.948 &  1.049 & -0.21 &    44.2 &  -69.6 &   -5.9 &   4.47 &   8.99 &   0.27 & 0.34 &  272 &  4.28 &  2.71 & 0.18 & 0.80 & 0.01 & 0.00 &    \\
 T0885-00642 &  1.105 &  1.044 & -0.10 &   -54.7 &  -11.7 &  -13.5 &   7.22 &   9.57 &   0.35 & 0.14 &  322 &  4.03 &  1.73 & 0.94 & 0.06 & 0.00 & 0.00 &    \\
 T0888-00115 &  0.680 &  1.016 & -0.28 &    -6.2 &   -3.6 &    5.9 &   8.40 &   8.72 &   0.33 & 0.02 &  283 &  3.94 &  2.37 & 0.97 & 0.03 & 0.00 & 0.00 &    \\
 T0888-00875 &  1.427 &  0.983 & -0.20 &    18.3 &   -1.3 &  -33.6 &   7.89 &   9.61 &   0.45 & 0.10 &  210 &  8.33 &  2.11 & 0.93 & 0.07 & 0.00 & 0.00 &    \\
 T0889-01220 &  0.737 &  0.995 & -0.30 &   -66.1 &  -34.2 &    7.5 &   6.04 &   9.38 &   0.30 & 0.22 &  223 &  4.47 &  2.51 & 0.55 & 0.12 & 0.33 & 0.00 &    \\
 T0897-00666 &  0.679 &  0.955 & -0.44 &   -78.2 &  -31.4 &    1.3 &   5.98 &   9.81 &   0.24 & 0.24 &  205 &  4.25 &  2.11 & 0.63 & 0.15 & 0.21 & 0.00 &    \\
 T0897-00860 &  1.278 &  1.053 & -0.01 &     5.7 &  -10.7 &   -5.4 &   7.80 &   8.75 &   0.38 & 0.06 &  378 &  3.03 &  1.16 & 0.97 & 0.03 & 0.00 & 0.00 &    \\
 T1442-00319 &  0.902 &  0.946 & -0.64 &    49.1 &    5.2 &  -22.5 &   7.40 &  11.23 &   0.54 & 0.21 &  397 &  5.93 &  2.88 & 0.94 & 0.06 & 0.00 & 0.00 & b   \\
 T1442-00453 &  1.404 &  0.941 & -0.56 &    27.8 & -104.7 &  -28.4 &   3.17 &   8.71 &   0.39 & 0.47 &  247 &  7.23 &  2.96 & 0.00 & 0.99 & 0.00 & 0.01 &  \\
 ..................... &  ......  & ......        &......  &......       &......      &......        &......  &......       &......    &......   &......   &......   &......      &......  &......  &......  &......  &  \\
 ..................... &  ......  & ......        &......  &......       &......      &......        &......  &......       &......    &......   &......   &......   &......      &......  &......  &......  &......  &  \\
\hline
\end{tabular}
\end{table}
\end{landscape}

\section{The thin disk traced by clump giants}
\label{s:clump}
Among the many studies that can be done with the new sample presented here, we focus on the properties of the
thin disk probed for the first time up to large distances above the Galactic plane, from a complete stellar 
sample and
with 3D kinematics and spectroscopic metallicities.
In order to work with a homogeneous sample, with well defined boundaries in both its local and distant 
counterparts,
we have selected clump giants on the basis of a colour and absolute magnitude
restriction : $0.9 \le B-V \le 1.1$, $0.0 \le  M_{\rm v} \le 1.6$.  According to Koen \& Lombard (2003),
this ensures the lowest contamination by other giants. Rejecting suspected binaries,
597 stars fall into these limits.  We further restrict the sample to the 396 stars having a 
probability higher than 80\% to belong to the thin disk.
In this section we investigate some basic distributions of this sample. 

\subsection{Raw metallicity distribution and vertical gradient}
\label{s:grad}
\begin{figure}[htbp]
\center
\caption{Metallicity distribution of thin disk clump giants
of the local (filled) and distant samples (red line). }
\includegraphics[width=8cm]{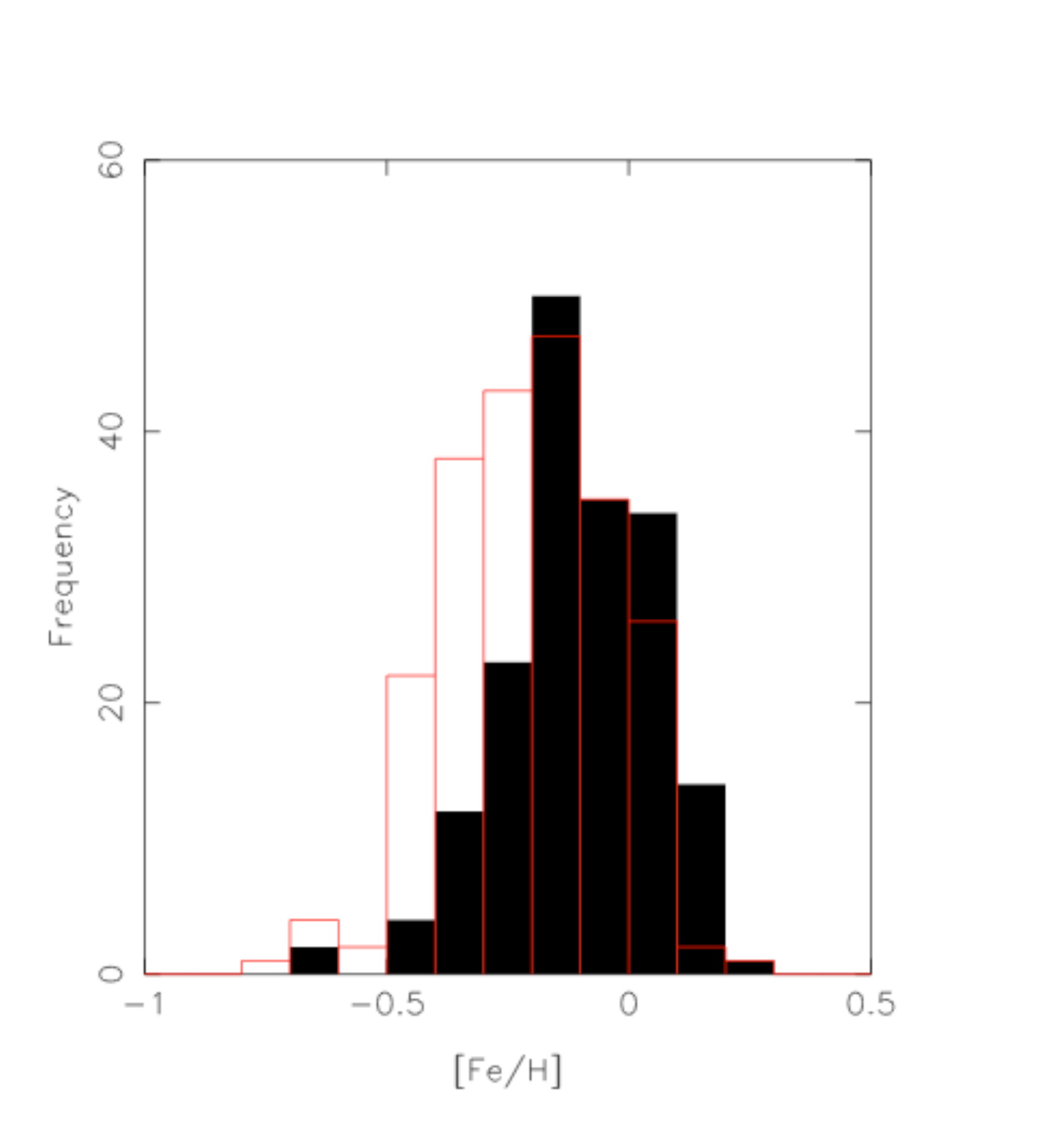}
\label{f:feh_histo_thin}
\end{figure}

\begin{figure}[htbp]
\center
\caption{Vertical gradient in the metallicity distribution of thin disk clump giants. }
\includegraphics[width=8cm]{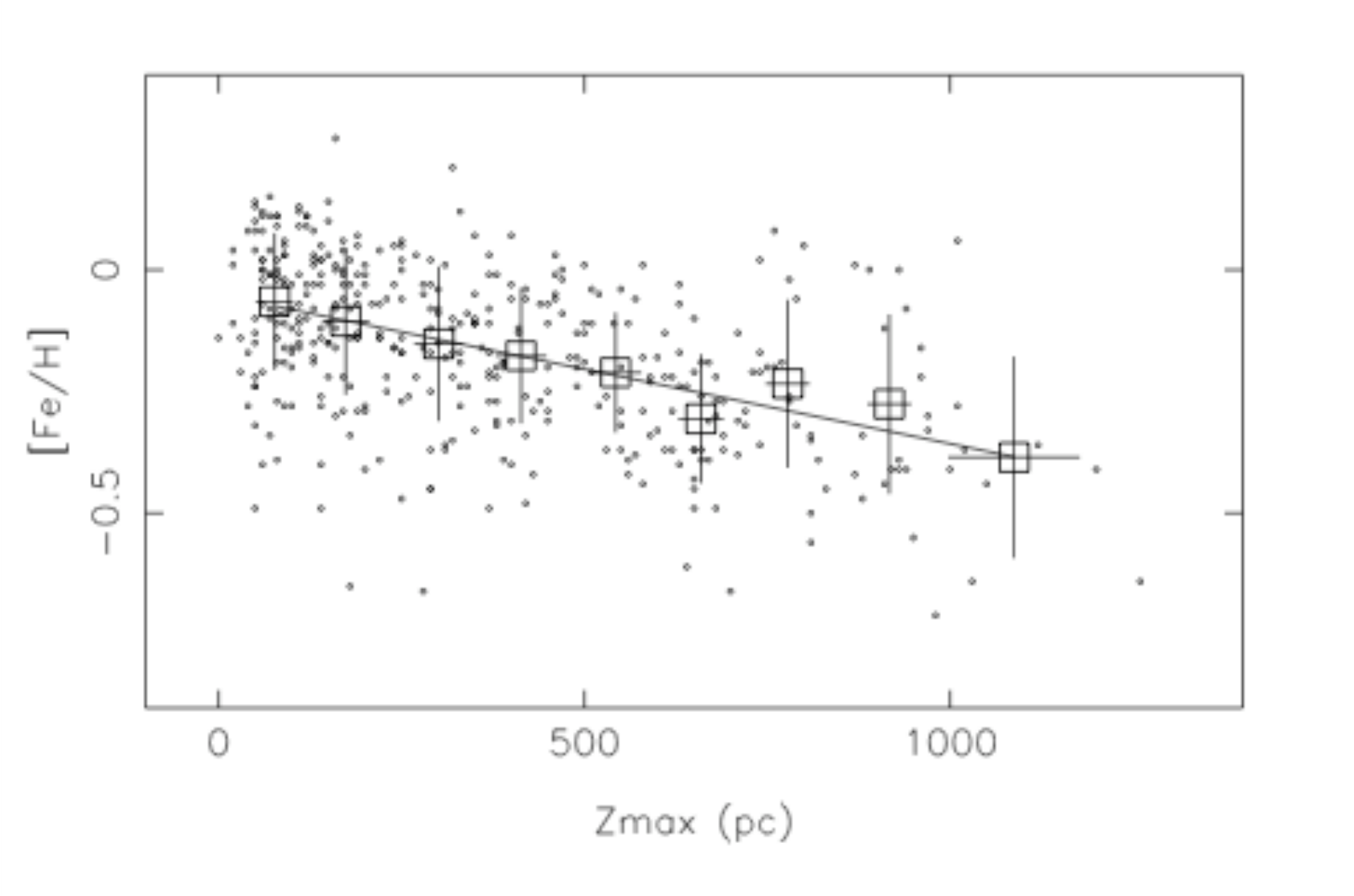}
\label{f:feh_grad_thin}
\end{figure}

We compare the metallicity distributions of the local and distant clump giants in Fig. 
\ref{f:feh_histo_thin}. 
The local sample has an average of [Fe/H]=-0.11 and a standard deviation of $\sigma_{[Fe/H]}=0.15$
whereas the distant sample has an average of [Fe/H]=-0.21 and a standard deviation of 
$\sigma_{[Fe/H]}=0.17$. The metallicity distribution of the thin disk is thus significantly shifted 
towards lower values at larger distance above the galactic plane. This is not due to the comparison 
of metallicities
coming from the literature for the local sample and from TGMET for the distant sample since we 
have verified that the two scales are consistent (Sect. \ref{s:TGMET}). More likely this difference 
indicates a vertical metallicity
gradient which is represented in Fig. \ref{f:feh_grad_thin}, using as the distance the maximum height 
from the plane, Zmax, reached by the star in its galactic orbit.
 A linear 
fit indicates a gradient of $\partial \mathrm{[Fe/H]} / \partial Z = -0.31 \pm 0.03$ dex kpc$^{-1}$. Taking for each star its current distance from the plane, instead of Zmax, leads to a consistent result of $\partial \rm{[Fe/H]} \partial z = -0.30 \pm 0.03$ dex kpc$^{-1}$.

 According to numerous previous studies, the existence of a vertical metallicity gradient in the thin disk seems to be firmly established. However the value of its amplitude, constrained by the observation of different kinds of tracers, at various distances from the Sun, is still oscillating between $\sim$ -0.25 and -0.35 dex kpc$^{-1}$. Using open clusters, Piatti et al. (\cite{pia95}) find -0.34 dex kpc$^{-1}$ whereas Carraro et al. (\cite{car98}) measure -0.25 dex kpc$^{-1}$ and Chen et al. (\cite{chen03}) measure $-0.295 \pm 0.050$ dex kpc$^{-1}$. Other studies are based, like ours, on field stars and have used kinematical information to select thin disc stars.  This is the case of Marsakov \& Borkova (\cite{mar06}) who have selected the most probable thin disk stars in their compilation of spectroscopic abundances, using their 3D velocties and orbital parameters. They measure a gradient of -0.29 $\pm$ 0.06 dex kpc$^{-1}$.  Barta\v{s}i\={u}t\.{e} et al. (\cite{bar03}) have observed 650 stars at high galactic latitude, up to 1.1 kpc, and identified thin and thick disk stars on the basis of their rotational lag. They measure a gradient of -0.23 $\pm$ 0.04 dex kpc$^{-1}$ in the thin disk. 

 The direct comparison of the metallicity distribution of our sample with other distributions probing different galactic volumes would imply a scaleheight correction. The reason is that metal-poor stars, which have hotter kinematics, have a larger scaleheight than more metal-rich stars, and may be under-represented in local samples. A correction,  relying on a mass model of the disk, would thus increase the number of metal-poor stars with hotter kinematics which would have been missed in our sample. On the contrary, stars more metal-rich than the Sun are supposed to be over-represented in local samples (see for instance Fig. 3 in Haywood \cite{hay06}). We have not attempted to correct such bias in our sample and we restrict  here the discussion to a qualitative comparison between dwarfs and giants.

When we compare the metallicity distribution of clump giants to that of dwarfs, as presented by  
Haywood (\cite{hay02}), we find a good agreement for the metal-poor side. 
We confirm with this new sample Haywood's finding that the thin disk is not an important contributor 
to stars with [Fe/H] 
$<$ -0.5. We find that 2.5\% of our sample has [Fe/H] 
$<$ -0.5 with the most metal-poor thin disk giant at [Fe/H]=-0.71. According to Fig. 3 in Haywood (\cite{hay06}), the scaleheight correction factor is comprised between $\sim$ 1.5 and 3.5 in the metallicity range -0.70 $<$ [Fe/H] $<$ -0.50. Taking this correction into account would not change substantially our findings.

On the contrary, we find a significant difference between clump giants and dwarfs for the metal-rich side
of the [Fe/H] histogram.
Haywood (\cite{hay02}) finds that 40-50\% of long-lived dwarfs have a metallicity higher than [Fe/H]=0 
whereas the proportion is only 20\%
in
our local sample and 13\% in our distant sample. Super Metal-Rich ([Fe/H] $>$ +0.20) FGK dwarfs  are quite usual 
in the Solar 
Neighbourhood whereas we have only two thin disk clump giants at [Fe/H]=+0.21 and [Fe/H]=+0.27. Our first guess 
was that such 
a low ratio of metal-poor stars and metal-rich stars 
was correlated with the colour cuts that we used to restrict the sample to clump giants. We have verified that this is not the case by comparing the metallicity
histograms of local giants ($0 \le  M_{\rm v} \le 1.6$) in the B-V intervals [0.9; 1.1] and [0.7; 1.2]. We found that the metal-poor
sides are similar. The ratio of metal-rich stars turns out to be slightly higher in the extended colour 
interval : 24\% instead of 20\%. We conclude that our adopted colour cutoff affects the metallicity distribution in 
a way that metal-rich stars are slightly under-represented. This bias is however not sufficient to reconcile the 
metallicity distribution of clump giants with that of dwarfs.

Pasquini et al. (\cite{pas07}) have also noticed a difference in the metallicity distribution of giants and 
dwarfs hosting planets. They propose as an interpretation the pollution of stellar atmospheres, causing a metal 
excess visible in the thin atmosphere of dwarfs, while diluted in the extended envelope of giants. Our sample suggests
that the difference is not limited to stars hosting planets so that the pollution hypothesis should be
investigated in a more general context. If validated in the general case, it would imply that dwarfs are not
appropriate to probe the chemical history of the Galaxy.

\subsection{Age - metallicity relation}
\label{s:AMR}
\begin{figure}[htbp]
\center
\caption{Age - metallicity relation of thin disk clump giants. The error bars represent the dispersion in each bin, including observational errors and cosmic scatter.}
\includegraphics[width=8cm]{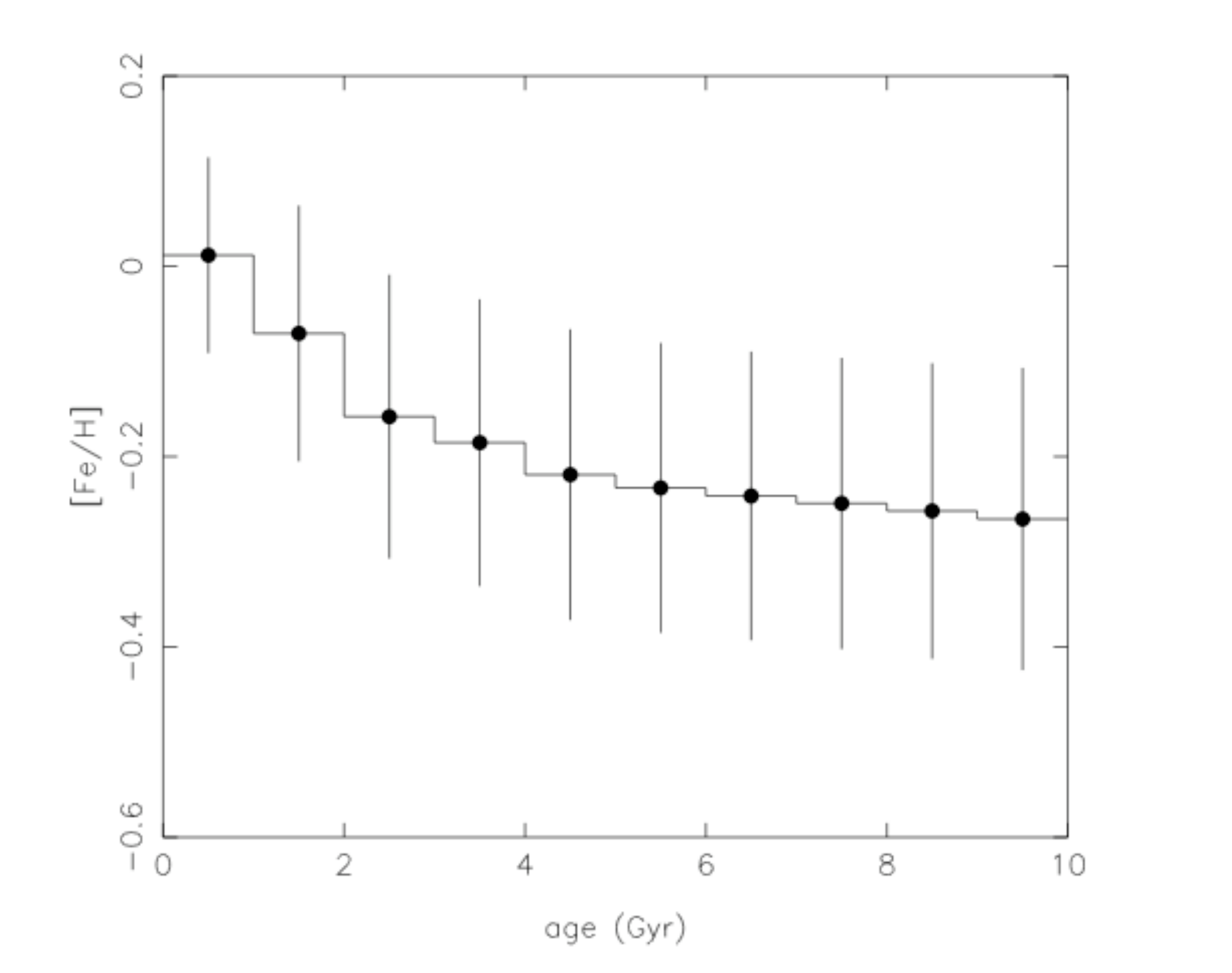}
\label{f:AMR}
\end{figure}

We use the same method as da Silva et al. (\cite{dasil06}) to determine the AMR of the sample.
For each time interval ($\Delta_t$=1 Gyr), we measure the cumulative [Fe/H] by adding the 
measured [Fe/H] of each star weighted by its PDF. With such a method, a given star contributes to
several bins which are consequently not independant. However it is a good way to take errors 
on ages into 
account. The mean metallicity and dispersion per age bin are presented in Fig. \ref{f:AMR} and
Table \ref{t:AMR}. A remarkable result is the low dispersion obtained at all 
ages. Subtracting the estimated observational error (0.09 dex for local stars, 0.13 dex for distant stars) yields to a cosmic scatter in
[Fe/H] lower than 0.12 dex. A transition occurs around 4 Gyr in both the mean metallicity and dispersion. From 10 Gyr to 4 Gyr, we
see a very smooth and regular increase of the mean metallicity, 0.01 dex per Gyr, 
with constant spread, which characterizes an homogeneous interstellar medium. An upturn occurs at 4 Gyr with
a steeper metallicity rise at younger ages.

 What is the state-of-the-art of the AMR in the Solar Neighbourhood and how does our relation compare with previous ones ?
 Despite numerous studies on the subject over nearly 30 years, there is still no consensus on the existence or not of a slope in the AMR, neither in the amplitude of the cosmic scatter. Major contributions like Edvardsson et al. (\cite{edv93}), Feltzing et al. (\cite{fel01}) and
 Nordstr\"{o}m et al. (\cite{nor04}), using classical isochrone ages, find little evidence for a slope in the relation of [Fe/H] with age, and a broad dispersion ($\sigma_{\rm [Fe/H]} > 0.20$ dex). In contrast,
 Rocha-Pinto et al. (\cite{roc00}),  using chromospheric ages, find a significant trend in the AMR, with lower dispersion ($\sigma_{\rm [Fe/H]} \sim 0.12$ dex). Pont \& Eyer (\cite{pon04}) have re-analysed Edvardsson et al.'s sample with a Bayesian approach and also find a significant trend with a dispersion $\sigma_{\rm [Fe/H]} < 0.15$ dex. We note that all these studies involve nearby dwarf stars. To our knowledge, the only AMR based on giants is that of da Silva et al. (\cite{dasil06}). Despite the small size of their sample, they find  like us metallicities rising from [Fe/H] $\sim$ -0.23 at  10.5 Gyr to [Fe/H] $\sim$ 0.00 at 0.5 Gyr. The shape of their AMR is however different of ours, shallower at young ages and steeper at old ages. The dispersion of their AMR is also much larger than ours, reaching 0.30 dex in the oldest age bins.
 
 We notice that the rather large metallicity variation that we observe in the 4 youngest bins in our AMR is also visible in the AMR derived by Nordstr\"om et
al. (\cite{nor04}) and by Feltzing et al. (\cite{fel01}). Both studies interpret this feature as a bias against young metal-poor dwarfs due to a colour cut.
This explaination is not valid for our sample since we have verified that our colour cuts only affect 
very slightly
the metal-rich part of the metallicity distribution (see previous Sect.). We thus conclude that this peculiar 
shape of the AMR is real.

 Piatti et al. (\cite{pia95}) and Carraro et al. (\cite{car98}) have corrected their AMR from the positional dependency, justified by the use of open clusters. Open clusters have a wide spatial distribution and trace different histories of the chemical evolution, depending on their galactocentric distances. The AMR has thus to be corrected from the observed radial metallicity gradient, which has an amplitude of 0.07 dex kpc$^{-1}$ according to Piatti et al. (\cite{pia95}), or 0.09 dex kpc$^{-1}$ according to Carraro et al. (\cite{car98}). Field stars are also supposed to be affected by a radial metallicity gradient. A consequence of orbital diffusion is that samples of nearby stars may include stars born in the inner or outer parts of the disk where the chemical enrichement may have been different from that of the Solar Neighbourhood. Such stars are easily recognized with their orbital parameters Rmin and Rmax, respectively perigalactic radius and apogalactic radius, different from that of the true local stars. It is worth mentioning that Edvardsson et al. (\cite{edv93}) have studied the AMR for stars restricted to the solar circle and still found a large and significant scatter. Our sample of thin disk clump giants is free from the influence of stars from other galactocentric distances since our kinematical selection has naturally eliminated stars on eccentric orbits. 

The question whether the AMR should be corrected from the vertical metallicity gradient is more difficult to assess. We note that Carraro et al. (\cite{car98}) have not attempted to correct their open cluster AMR from the observed vertical metallicity gradient. Moreover they
argue that "In the case of field stars, orbital diffusion is expected to be effective enough to smooth out a vertical metallicity gradient within a single-age population, so that the vertical structure of the disk is dominated by the different scaleheights of different age populations". We also note that, in the case of field star AMRs, while the radial migration is often refered to (Edvardsson et al. \cite{edv93}, Haywood \cite{hay06}), the influence of the vertical metallicity gradient is not discussed.

\begin{table}[t]
\caption{
\label{t:AMR}
Age-metallicity relation derived from our sample of thin disk clump giants. N is the number of 
stars contributing to each age bin. It is fractional because we use the complete probability 
function of each star to bin the age axis (see text).}
\begin{center}
\begin{tabular}{c c c c}
\hline
\hline
$<{\rm t}>$ (Gyr) & $<{\rm [Fe/H]}>$ & $\sigma_{\rm [Fe/H]}$ & $<{\rm N}>$  \\
\hline
0.5 & +0.01 & 0.10 & 43.8 \\
1.5 & -0.07 & 0.13 & 79.3 \\
2.5 & -0.16 & 0.15 & 54.6 \\
3.5 & -0.19 & 0.15 & 41.9 \\
4.5 & -0.22 & 0.15 & 34.1 \\
5.5 & -0.23 & 0.15 & 28.1 \\
6.5 & -0.24 & 0.15 & 28.5 \\
7.5 & -0.25 & 0.15 & 22.2 \\
8.5 & -0.26 & 0.16 & 15.3 \\
9.5 & -0.27 & 0.16 & 15.7 \\
\hline
\end{tabular}
\end{center}
\end{table} 

\subsection{Age - velocity relation}
\label{s:AVR}
\begin{figure}[htbp]
\center
\caption{Age - velocity relations of distant clump giants, the most probable thick disk, Hercules stream and 
halo stars being excluded. }
\includegraphics[width=8cm]{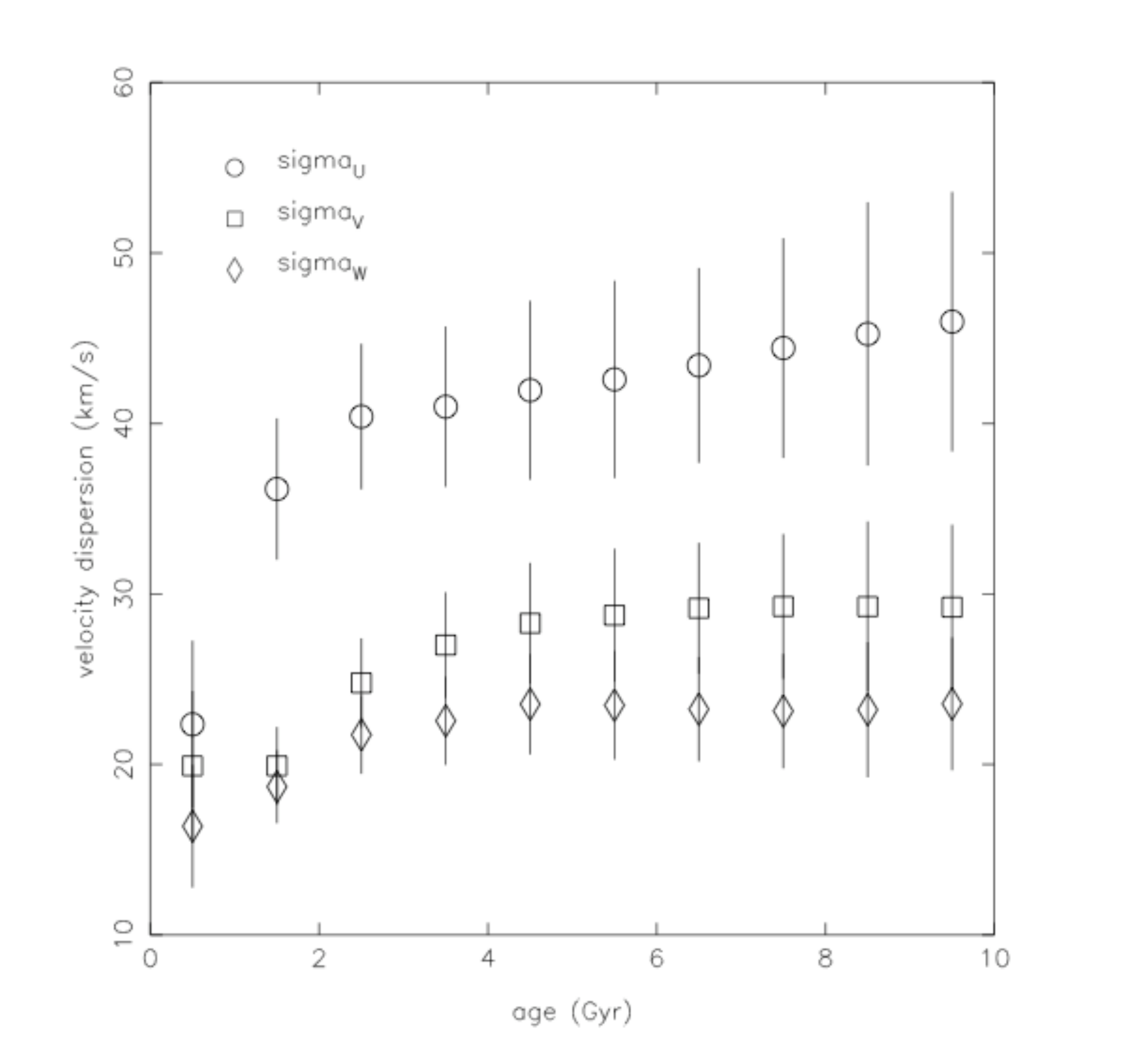}
\label{f:AVR}
\end{figure}

The thin disk AVR has been revisited recently by Seabroke \& Gilmore (\cite{sea07})
using the data of Nordstr\"om et al. (\cite{nor04}) and Famaey et al. (\cite{fam05}). They show that the 
kinematical streams in these local samples do not permit one to safely constrain the relations in the
U and $V$ directions, contrary to the $W$ direction where the samples are well mixed. Our sample of
clump giants, spanning larger distances from the Galactic plane, is well suited to investigate
these relations. However, for such a purpose, we cannot work on the thin disk sample which was built to study
the metallicity and age distributions, in Sections
\ref{s:grad} and \ref{s:AMR}.
Our selection of thin disk stars on a kinematical criterion has favoured stars in the central parts of
the velocity ellipsoid, with moderate velocities, resulting in a serious kinematical 
bias. In order to study
how the velocity dispersions increase with time, we need to work also with the warmer part of
the thin disk, but excluding as well as possible stars which do not follow the kinematical 
behaviour of the thin disk. To do so, we consider our distant sample of clump giants and reject stars
having a probability higher than 80\% to belong to the thick disk, the Hercules stream and the halo,
resulting in 320 stars. Results are presented in Fig. \ref{f:AVR} and Table \ref{t:AVR}. 

An important question
is whether the dispersions saturate at a given age of the thin disk. Seabroke \& Gilmore (\cite{sea07}) have
shown that local data are in agreement with several models of disk heating : continuous or with saturation at
4.5, 5.5 and 6.5 Gyr. Our data show evidence for a transition at $\sim$ 5 Gyr, with saturation occuring in 
V at 29 \kms and in $W$ at 24 \kms. The velocity dispersion in $U$ seems to increase smoothly, reaching 46 \kms at 9.5 Gyr. 
A consequence is that the 
velocity ellipsoid axis ratios $\sigma_V / \sigma_U$ and $\sigma_W / \sigma_U$ are not constant. The ratio 
$\sigma_V / \sigma_U$ is related to the Oort constants and is expected to be $\sim$0.5. Here this ratio is varying from
0.55 at 1-2 Gyr to a maximum value of 0.68 at 4-5  Gyr. The ratio 
$\sigma_W / \sigma_U$ is related to the scattering process responsible of the dynamical heating of the disk. 
With our data, it has a maximum value
of 0.56 at 4-5 Gyr. Although these ratios are supposed to be constant in an axisymmetric Galaxy, there are previous
reports of variations related to colour or spectral types (e.g. Mignard \cite{mi00}) .

 We mention the study by Vallenari et al. (\cite{val06}) who have also probed the thin disk kinematics towards the NGP. Their method is however significantly different from ours since they analyse, through a galactic  model, proper motions and the colour magnitude diagram  of $\sim$ 15000 stars down to V=20. Their best-fit for the velocity dispersions of the thin disk, presented in 4 age bins, differ significantly from ours, especially in the oldest age bin (7-10 Gyr) where their values are lower by 3$\sigma$.

Simple statistics on our
sample gives $(\sigma_U,\sigma_V,\sigma_W)=(41.5, 26.4, 22.1)$\kms, significantly higher than
values determined from late-type Hipparcos stars (e.g. Bienaym\'e \cite{bien99}, Mignard \cite{mi00}). Although 
we cannot 
rule out the contamination of the sample with thick disk stars, it nicely compares to recent results by de Souza 
\& Teixeira (\cite{desou07}) who show that Mignard's sample is better explained  by the superposition of
2 velocity ellipsoids the hotter one having $(\sigma_U,\sigma_V,\sigma_W)=(41.0, 27.0, 19.0)$\kms.
It is also worth noticing in Table \ref{t:AVR} that the mean $U$ and $W$ are roughly constant at all ages 
whereas $V$ declines
from $\sim$ -14 \kms to -21 \kms. We retrieve for $U$ and $V$ the Solar motion with respect to late-type stars,
as determined by Mignard (\cite{mi00}), although we find a significant difference in $W$. We get a mean value of
$W_{\odot}=11.5$ \kms, whereas he finds values around 7 \kms. We recall that our $W$ velocities of the distant 
stars
at the NGP rely mainly on radial velocities, which have an accuracy better than 1 \kms, and thus are not affected 
by uncertainties on distances and proper motions.

\begin{table}[t]
\caption{
\label{t:AVR}
Age-velocity relation derived from 320 distant clump giants, the most probable thick disk, Hercules stream and 
halo stars being excluded.}
\begin{center}
\begin{tabular}{c c c c c c c c}
\hline
\hline
t (Gyr) & $U$ & $\sigma_U$ & $V$ & $\sigma_V$ & $W$ & $\sigma_W$ & N \\
\hline
 0.50 & -11.3 &  22.3 & -14.2 &  19.9 & -14.2 &  16.4 & 10.3 \\
 1.50 & -11.6 &  36.2 & -13.9 &  19.9 & -10.6 &  18.7 & 37.9 \\
 2.50 &  -9.6 &  40.4 & -15.5 &  24.8 & -10.6 &  21.7 & 44.7 \\
 3.50 &  -9.3 &  41.0 & -16.7 &  27.0 & -10.7 &  22.6 & 37.8 \\
 4.50 &  -9.2 &  41.9 & -17.5 &  28.3 & -11.1 &  23.5 & 31.9 \\
 5.50 &  -9.3 &  42.6 & -17.9 &  28.8 & -11.6 &  23.5 & 27.0 \\
 6.50 &  -9.8 &  43.4 & -18.6 &  29.2 & -12.2 &  23.2 & 28.8 \\
 7.50 &  -9.7 &  44.4 & -19.5 &  29.3 & -12.8 &  23.1 & 23.7 \\
 8.50 &  -9.2 &  45.3 & -20.3 &  29.3 & -13.1 &  23.2 & 17.1 \\
 9.50 &  -8.7 &  46.0 & -21.0 &  29.2 & -13.1 &  23.6 & 18.2 \\
\hline
\end{tabular}
\end{center}
\end{table}

%
\section{Summary}
The data presented here are the result of several years of effort to obtain high resolution
spectra for a large and complete sample of clump giants. Besides our own observations, on
the ELODIE spectrograph at OHP, 
we have also taken advantage of other available material like the Hipparcos and Tycho2
catalogues, the [Fe/H] catalogue (Cayrel de Strobel et al. \cite{cay01}) updated with a number 
of new references and the PARAM code to derive ages (da Silva et al. \cite{dasil06}). We have described
 how these data were combined to provide a catalogue of stellar parameters for
891 stars, mainly giants, giving atmospheric parameters with spectroscopic metallicities, absolute magnitudes 
and distances, 
galactic velocities $(U,V,W)$, orbits, ages and population membership probabilities.

Our main motivation in conducting this project was to probe the Galactic disk using an 
unbiased and significant sample, with high quality data, in particular with spectroscopic metallicities 
and accurate distances and 
radial velocities. We have chosen to observe giants in the direction of the NGP in order to reach 
distances
to the galactic plane, up to 1 kpc, which are not covered by spectroscopic surveys, usually limited to the closer
Solar Neighbourhood. Clump giants are particularly well suited for this purpose.  Compared to previous studies on the subject, our analysis presents several improvements, which are briefly outlined:

\begin{itemize}
\item for binning the age axis, we have considered for each star its entire age PDF, instead of averaging it, following da Silva et al. (\cite{dasil06})
\item we have considered several kinematical populations likely to be present in our sample: the thin disk, the thick disk, the Hercules stream and the halo 
\item in order to study the thin disk metallicity and age distributions, we have taken care to select stars with the highest probability of belonging to this population
\item in order to study the thin disk velocity distribution, we have taken care to reject the most probable non thin disk stars
\end{itemize}

Our results  are summarized as follows :

\begin{itemize}
\item we do not find any young metal-poor stars nor old metal-rich stars, contrary to common findings in dwarf samples
\item the old thin disk and the thick disk overlap in the metallicity range -0.70 $\leq$ [Fe/H] $\leq$ -0.30 and age range 8-10 Gyr
\item among stars with accurate individual ages, we do not find any young thick disk stars 
\item the metallicity distribution of our sample of thin disk clump giants extends down to [Fe/H]$\simeq$-0.70, but the fraction of stars with [Fe/H]$\leq $-0.50 is only 2.5\%
\item the metallicity distributions of giants and dwarfs differ significantly on the metal-rich side: 
metal-rich giants are less frequent
\item a vertical metallicity gradient is measured in the thin disk: $\partial \mathrm{[Fe/H]} / \partial Z = -0.31 \pm 0.03$ dex kpc$^{-1}$
\item the AMR of the thin disk presents a low dispersion, implying a cosmic scatter lower than 0.12 dex, in agreement with previous findings by Rocha-Pinto et al. (\cite{roc00}) and Pont \& Eyer (\cite{pon04})
\item 2 regimes are visible in the AMR of the thin disk : from 10 Gyr to 4 Gyr, the metallicity increases smoothly by 0.01 dex per Gyr, while for younger stars the rise of [Fe/H] is steeper
\item in the thin disk, the $V$ and $W$ dispersions saturate at 29 and 24 \kms respectively at $\sim$ 4-5 Gyr, whereas $U$ shows 
continuous heating
\item the Solar motion is found to be nearly constant in $U$ and $W$ with respect to stars of all ages, while the amplitude of the asymmetric drift increases from 14 to 21 \kms with respect to young and old stars respectively
 \end{itemize}

%
\begin{acknowledgements}
We are grateful to L. Girardi for computing the ages for the 891 stars of this sample.
This  research  has made  use  of  the  SIMBAD and  VIZIER  databases,
operated at CDS, Strasbourg,  France. It is
based  on data from  the ESA {\it  Hipparcos} satellite
(Hipparcos and Tycho2 catalogues). 
\end{acknowledgements}


\begin{thebibliography}{}

\bibitem[1993]{allen}
Allen, C., \& Santillan, A. 1993, RMxAA, 25, 39

\bibitem[2001]{alo01}
Alonso, A., Arribas, S., \& Mart\' inez-Roger, C. 2001, A\&A, 376, 1039

\bibitem[1996]{bar96}
Baranne, A., Queloz, D., Mayor, M. et al. 1996, A\&AS, 119, 373

\bibitem[2000]{bar00}
Barbier-Brossat, M. \& Figon, P. 2000, A\&A Sup., 142, 217

\bibitem[2003]{bar03}
Barta\v{s}i\={u}t\.{e}, S., Aslan, Z., Boyle, R. P., Kharchenko, N. V., Ossipkov, L. P., \& Sperauskas, J. 2003, Baltic Astron., 12, 539

\bibitem[2007]{bia07}
Biazzo, K., Pasquini, L., Girardi, L., et al. 2007, A\&A, 475, 981

\bibitem[1998]{blac98}
Blackwell, D.E., \& Lynas-Gray, A.E. 1998, A\&AS, 129, 505

\bibitem[1999]{bien99}
Bienaym\'e O. 1999 A\&A, 341, 86

\bibitem[2005]{bie05} 
Bienaym\'e, O., Soubiran, C., Mishenina, T.V., Kovtyukh, V.V., \& Siebert, A. 2005, A\&A, 456, 1109
(Paper III)

\bibitem[1998]{car98}
Carraro, G., Ng, Y. K., \& Portinari, L. 1998, MNRAS, 296, 1045

\bibitem[2001]{cay01}
Cayrel de Strobel, G., Soubiran, C., \& Ralite, N. 2001, A\&A, 373, 159

\bibitem[2003]{chen03}
Chen, L., Hou, J. L.,  \& Wang, J. J. 2003, AJ, 125, 1397

\bibitem[2006]{dasil06}
da Silva L., Girardi L., Pasquini L. et al 2006, A\&A, 458, 609

\bibitem[2007]{desou07}
de Souza, R.E. \& Teixeira, R. 2007, A\&A, 471, 475

\bibitem[1998]{diben98}
di Benedetto, G.P. 1998, A\&A, 339, 858

\bibitem[1993]{edv93}
Edvardsson, B., Andersen, J., Gustafsson, B., Lambert, D. L., Nissen, P. E., \& Tomkin, J. 1993, A\&A, 275, 101

\bibitem[1997]{esa97}
ESA 1997, The Hipparcos and Tycho Catalogues, (Noordwijk) Series: ESA-SP 1200

\bibitem[2005]{fam05}
Famaey, B., Jorissen, A., Luri, X. et al. 2005, A\&A, 430, 165

\bibitem[2001]{fel01}
Feltzing, S., Holmberg, J., \& Hurley, J. R. 2001, A\&A, 377, 911

\bibitem[2000]{gir00}
Girardi L., Bressan A., Bertelli G., \& Chiosi C. 2000, A\&AS, 141, 371

\bibitem[1998]{har98}
Harmanec, P. 1998, A\&A, 335, 173

\bibitem[2002]{hay02}
Haywood, M. 2002, MNRAS, 337, 151

\bibitem[2006]{hay06}
Haywood, M. 2006, MNRAS, 371, 1760

\bibitem[2000]{hog00} 
H{\o}g, E., Fabricius, C., Makarov, V.\,V. et al. 2000, A\&A, 363, 385

\bibitem[1987]{joh87}
Johnson, D.R.H. \& Soderblom, D.R. 1987, AJ, 93, 864

\bibitem[2005]{jor05}
J{\o}rgensen, B.R. \& Lindegren, L. 2005, A\&A, 436, 127

\bibitem[2003]{koen03}
Koen, C. \& Lombard, F. 2003, MNRAS, 343, 241

\bibitem[1998]{kat98} 
Katz, D., Soubiran, C., Cayrel, R. et al.  1998, A\&A, 338, 151

\bibitem[2006]{kov06}
Kovtyukh, V.V., Soubiran, C., Bienaym\'e, O., Mishenina, T.V., \& Belik, S. I. 2006,
MNRAS, 371, 879

\bibitem[2006]{mar06}
Marsakov, V.A. \& Borkova, T.V. 2006, AstL, 32, 376

\bibitem[1990]{mcw90}
McWilliam, A. 1990, ApJS, 74, 1075

\bibitem[2000]{mi00}
Mignard, F. 2000, A\&A, 354, 522

\bibitem[2006]{mish06}
Mishenina, T.V., Bienaym\'e, O., Gorbaneva, T.I., Soubiran, C., Charbonnel, C., 
Korotin, S.A., \& Kovtyukh, V.V. 2006, A\&A, 456, 1109

\bibitem[2004]{mou04}
Moultaka, J., Ilovaisky, S.A., Prugniel, P., \& Soubiran, C. 2004, PASP, 116,693

\bibitem[2004]{nor04}
Nordstr\"om B., Mayor, M., Andersen, J., et al. 2004, A\&A, 418, 989

\bibitem[2007]{pas07}
Pasquini, L., D\"ollinger, M.P., Weiss, A. et al. 2007, A\&A 473, 979

\bibitem[1995]{pia95}
Piatti, A., Claria, J. J.,  \& Abadi, M. G. 1995, AJ, 110, 2813

\bibitem[2004]{pon04}
Pont, F., \& Eyer, L. 2004, MNRAS, 351, 487

\bibitem[2004]{pru04}
Prugniel, P., \& Soubiran, C. 2004, astro-ph/0409214

\bibitem[2001]{pru01}
Prugniel, P., \& Soubiran, C. 2001, A\&A, 369, 1048

\bibitem[2005]{ram05}
Ram\'\i rez I., \& Mel\'endez J. 2005, ApJ, 626, 446

\bibitem[2000]{roc00}
Rocha-Pinto, H. J., Maciel, W. J., Scalo, J., \& Flynn, C. 2000, A\&A, 358, 850 

\bibitem[2007]{sea07}
Seabroke G.M., \& Gilmore G. 2007, MNRAS, 380, 1348

\bibitem[2003]{sie03}
Siebert, A., Bienaym\'e, O., \& Soubiran, C. 2003, A\&A, 399, 531 (Paper II)

\bibitem[2003]{sou03}
Soubiran, C., Bienaym\'e, O., \& Siebert, A. 2003, A\&A, 398, 141 (Paper I)

\bibitem[1998]{sou98} 
Soubiran, C., Katz, D., \& Cayrel, D. 1998, A\&A, 133, 221

\bibitem[2006]{val06}
Vallenari, A., Pasetto, S., Bertelli, G., Chiosi, C., Spagna, A., \& Lattanzi, M. 2006, A\&A, 451, 125
\end{thebibliography}
\end{document}